\documentclass[acmsmall,nonacm,screen]{acmart}
\setcopyright{none}
\renewcommand\footnotetextcopyrightpermission[1]{}
\usepackage{booktabs}      % nice tables
\usepackage{amsthm}
\usepackage{mathpartir}    % inference rules (optional; comment out if unused)
\usepackage{listings}      % code listings (or use lstlisting/minted)
\usepackage{mathpartir,graphicx}
\usepackage{tikz}\usetikzlibrary{arrows.meta,calc}

\title{Program Analysis with Prophecy and History Variables in the Nexis Compiler}
\author{Martin Rinard}
\affiliation{%
  \institution{National University of Singapore and Massachusetts Institute of Technology}
  \city{Singapore and Cambridge, MA}
  \country{Singapore and USA}
}
\email{rinard@csail.mit.edu}

\begin{abstract}

We present prophecy variables for forward formulations of program
analysis problems that require information about the future 
execution of the program. We specify prophecy and history variables
via a domain specific language that augments 
the step rules of the base operational semantics with subset
inclusion constraints over the prophecy and history variables. 
This tight coupling between the prophecy and history variable specification
and the operational semantics promotes the construction
of correctness and optimality proofs for program transformations,
with the proofs structured as forward simulations between the original and
transformed versions of the program. 
In comparison with traditional dataflow approaches, this approach eliminates
mechanisms such as explicit control flow graphs, abstraction functions, 
concretization functions, Galois connections, and separate backward 
and forward analyses. 

We present a verified implementation of prophecy and history variables and
use the implementation to prove correctness and optimality properties
of two classic transformations, partial dead code elimination and
lazy code motion, that use both prophecy and history variables. 
To the best of our knowledge, these proofs are the first machine checked 
correctness and optimality proofs for these transformations. 

\end{abstract}

\begin{document}
\pagestyle{plain}

\maketitle

\section{Introduction}
\label{sec:intro}

The semantics of programming languages are often defined with an operational semantics
that specifies program executions as {\em traces} --- sequences of transitions over 
dynamic program states~\cite{Winskel93}. Program analyses, in contrast, are often specified 
over static program representations such as control flow graphs~\cite{DragonBook,appel2004modern,muchnick1997advanced,kennedy2001optimizing,cooper2011engineering}. Proofs of program analysis properties typically use mechanisms such as abstraction functions, 
concretization functions, and Galois connections to bridge the gap, with forward simulation
proofs mapping paths in the static program representation to traces in the
operational semantics~\cite{CousotC77,CousotC92}.

We propose an alternative approach that defines program analyses directly over the 
base operational semantics of the programming language. This approach augments
program execution states with {\em ghost variables} that contain the analysis result. 
% - use shadow variables in final version. closer match to dynamic analysis. 
It augments the transition rules of the base semantics with {\em ghost variable constraints} 
that define the analysis. By specifying the analysis directly over the 
operational semantics, this approach tightly couples the analysis with the semantics and 
promotes the construction of forward simulation proofs by eliminating
static program representations and associated machinery from the specification of the analysis
and proofs that use program analysis results. 

{\em History variables} that maintain information from the past execution of the program
take the place of forward dataflow analyses in traditional dataflow analysis frameworks.
At each point in the execution of the program this information is available in and
maintainable over the sequence of transitions that led to that point in the execution. 
Traditional backward dataflow analyses, however, extract information about the future
execution of the program, which is not available in the transition sequence until the
program actually performs that future execution. 

We therefore use {\em prophecy variables} to specify these kinds of analyses. Prophecy
variables initially arose in the context of state machine refinement proofs~\cite{AbadiL91}, 
which use forward simulation proofs over execution traces to prove that a concrete implementation 
refines an abstract specification. In this context prophecy variables enable forward
proofs by nondeterministically predicting future choices that are then checked 
when the information required to determine the choice becomes available~\cite{AbadiL91}. 
In our context prophecy variables predict analysis results that depend on the 
future execution of the program, with the predictions checked as the information
becomes available in the execution. This mechanism unifies forward and backward
analyses and promotes the construction of forward simulation proofs that
use information about the future execution of the program. 

\subsection{Prophecy and History Variable Specifications}

We present a domain specific language for specifying prophecy and history
variables. In this language, prophecy and history variables range over sets
of constructs from the analyzed program such as variables, expressions, and assignments. 
The values of prophecy and history variables are defined by augmenting the 
preconditions of the step rules in the operational semantics with 
subset inclusion constraints over the prophecy and history variables. 
For prophecy variables $\pi$, programs in the specification language 
augment step rules that define transitions $\langle n, \sigma \rangle \rightarrow \langle n', \sigma' \rangle$
in the base semantics with 1) nondeterministic $\mathsf{predict}$ clauses that predict prophecy
variable values $\pi(n')$ at $\langle n', \sigma' \rangle$ that will 2) satisfy future $\mathsf{check}$ clauses
(also specified at step rules in the augmented semantics). 
History variables $\eta$ have $\mathsf{update}$ clauses that specify the value $\eta(n')$ at $n'$. 
Taken together, the rules in the augmented operational semantics specify 
{\em valid} and {\em extremal} solutions to the 
$\mathsf{predict}$, $\mathsf{check}$, and $\mathsf{update}$ constraints ---
valid solutions satisfy the constraints; extremal solutions are the least
or greatest valid solutions (under the subset inclusion order). 

\subsection{Prophecy and History Variable Implementation}

We present an implementation of the domain specific language in the Nexis compiler system. 
Given a prophecy and history variable specification written in the
language, the Nexis implementation automatically generates an implementation of the
specification. When run on a program written in the base language, 
this implementation produces a bitvector representation of the
valid and extremal prophecy and history variable values. 
To compute these values, Nexis automatically generates forward
(for history variables) and backward (for prophecy variables) 
dataflow analysis problems and code that invokes a worklist
algorithm to solve these problems. We note that other solution
mechanisms, for example repeated program execution, may be appropriate
depending on the context~\cite{brahmakshatriya2026backwardsdataflowanalysisusing}.

In addition to the implementation, Nexis also generates proofs of the following
properties: 
\begin{itemize}
\item {\bf Progress:} Every transition in the base operational semantics
has a corresponding transition in the augmented operational semantics, with the
transitions preserving the values of the variables in the base operational semantics. 
A key property that this proof must establish is that the prophecy variable
predictions correctly predict the downstream prophecy variable checks (so that 
unsatisfied checks do not eliminate base semantics transitions). 
\item {\bf Preservation:} Every transition in the augmented operational
semantics has a corresponding transition in the base operational semantics. 
\item {\bf Validity:} The dataflow solver produces prophecy and history
variable values that satisfy the specified constraints. 
\item {\bf Extremality:} The dataflow solver produces prophecy and history
variable values that are the greatest or least solution to the 
specified constraints as appropriate. 
\end{itemize}
Together, progress and preservation establish a bisimulation between 
the base semantics and the augmented semantics. This bisimulation 
ensures that augmented semantics correctly preserves the base semantics and enables
correctness and optimality proofs that use the history and prophecy variables
as defined by the augmented semantics to reason 
about executions in the base semantics. 

Validity and extremality establish that the constraints in the step rules
hold for the prophecy and history variable values that Nexis
produces and that transforms work with. Transform correctness and
proofs are typically formulated as simulation proofs
over traces from the original and transformed programs. These 
simulation proofs establish invariants that hold at corresponding
points in the original and transformed traces, with the invariant
using the prophecy and history variables to state a refinement
mapping between program states at corresponding points.  
The inductive proof of the invariant pushes the invariant across matching transitions from the 
original and transformed programs, using the constraints from the 
step rules of the augmented semantics to reestablish the invariant. 

Defining the prophecy and history variables as step rule constraints
in the operational semantics therefore directly supports the proof 
structures that establish the desired transform correctness properties. 

% talk about what happens if predict is wrong and can't satisfy check - proof breaks

\subsection{Partial Dead Code Elimination and Lazy Code Motion} 

We present two program transformations, 
partial dead code elimination (PDCE~\cite{DBLP:conf/pldi/KnoopRS94,DBLP:journals/toplas/KnoopRS94}) 
and lazy code motion (LCM~\cite{DBLP:conf/pldi/KnoopRS92,DBLP:journals/toplas/KnoopRS94})
formulated with prophecy and history variables. PDCE has a prophecy variable and 
a history variable. LCM has six ghost variables (including multiple prophecy and
history variables). We prove correctness and optimality for both transformations. 
The proofs leverage the connection between the prophecy and history variables
and the operational semantics --- the correctness proofs 
are structured as forward simulation proofs over the operational 
semantics of the base language, with the ghost variables enabling the 
precise statement of refinement mappings that prove the correspondence
between the transformed and original programs at each transition. 

All of these proofs are stated over universally quantified valid
and/or extremal (as appropriate) values of the prophecy and 
history variables. In contrast to previous 
proofs, these proofs do not rely on any property of any dataflow
analysis. The prophecy and history variable specifications therefore
provide a clean abstraction boundary, with the dataflow analysis
relegated to an implementation mechanism behind the abstraction. 

The analyses, transformations, correctness proofs, and optimality proofs are all formalized and
machine checked in Lean 4. To the best of our knowledge, these are
the first machine checked correctness and optimality proofs of
any kind for these transformations and the first to ground the 
correctness and optimality on the operational semantics
of the underlying base language (as opposed to reasoning about
paths in a control flow graph).

\subsection{Contributions}

This paper makes the following contributions:
\begin{itemize}

\item {\bf Prophecy Variables for Program Analysis:} It formalizes prophecy variables for 
forward formulations of program analyses that work with information
about the future execution of the program. This formalization 
specifies prophecy (and history) variables via an augmented operational
semantics, closely coupling the analysis with the operational semantics
to support forward simulation proofs. 

\item {\bf Dataflow Analysis Relegation:} Valid and extremal solutions
to the prophecy and history variable constraints provide a declarative
program analysis abstraction, relegating dataflow analysis and its
associated machinery to an implementation mechanism isolated behind the abstraction, 
with the dataflow implementation verified to generate correct valid and
extremal solutions to the constraints that define the values of the prophecy and
history variables. 

\item {\bf Analyses and Transformations:} It formulates 
partial dead code elimination and lazy code motion as prophecy and
history variable driven analyses and transformations. It uses
these formulations to drive correctness and optimality proofs, 
each universally quantified over arbitrary valid and extremal 
solutions to the prophecy and history variable constraints. 

\item {\bf Machine Checked Proofs:} It presents the first
machine checked correctness and optimality proofs for partial
dead code elimination and lazy code motion, including the
first correctness and optimality proofs of any kind for a variant of lazy code motion that
uses edge insertions of moved expression evaluations to avoid a
critical edge splitting preprocessing pass. These proofs leverage
the Lean 4 prophecy and history variable formalization and are
fully machine checked. 

\item {\bf Implementation:} The entire system ---
prophecy and history variable infrastructure, language definitions, 
operational semantics, proofs, and a verified source to assembly compiler ---
was implemented by a coding agent (Claude Code Opus 4.8) generating Lean 
4 code supervised by the author in VS Code. 

\end{itemize}

The remainder of the paper is structured as follows. Section~\ref{sec:base-language} presents
the base language and its operational semantics and evaluation rules. Sections~\ref{sec:pdce} 
and \ref{sec:lcm} present the prophecy and history variable implementations
of partial dead code elimination and lazy code motion. Section~\ref{sec:nexis} presents the 
Nexis implementation including the prophecy and history variable domain specific language
and its implementation. Section~\ref{sec:related} presents related work. We conclude
in Section~\ref{sec:conclusion}.

This paper comes with a repository containing the full Nexis system --- verified compiler,
prophecy and history variable domain specific language implementation, PDCE and
LCM implementations with correctness and optimality proofs, and test programs. 
Section~\ref{sec:roadmap} provides a roadmap to this repository. 

% ===== Fig. 0a — Base language syntax (three-address CFG IR) =====
\begin{figure}[thbp]
\small
\noindent$
\begin{array}{@{}l@{\ \ }r@{\ }c@{\ }l@{}}
\text{values}    & v      & \in & \mathsf{Val}=\mathsf{Int64}\\
\text{nodes}     & n,n'   & \in & \mathsf{Node}=\mathbb{N}\\
\text{variables} & x,y    & ::= & \mathsf{orig}\,s \mid \mathsf{tmp}\,i \quad (s\in\mathsf{String},\ i\in\mathbb{N})\\
\text{atoms}     & a,b    & ::= & \mathsf{var}\,x \mid \mathsf{imm}\,v\quad(\text{a variable, or an immediate value})\\
\text{unops}     & \odot  & ::= & \mathsf{neg} \mid \mathsf{not}\\
\text{binops}    & \oplus & ::= & \mathsf{add} \mid \mathsf{sub} \mid \mathsf{mul} \mid \mathsf{div} \mid \mathsf{mod} \mid \mathsf{and} \mid \mathsf{or} \mid \mathsf{xor} \mid \mathsf{shl} \mid \mathsf{lshr} \mid \mathsf{ashr}\\
                 &        &     & \mid\ \mathsf{eq} \mid \mathsf{ne} \mid \mathsf{lt} \mid \mathsf{le} \mid \mathsf{ltu} \mid \mathsf{leu}\\
\text{exprs}     & e      & ::= & a \mid \odot\,a \mid a \oplus b\\
\text{commands}  & \iota  & ::= & (x := e)\to n \ \mid\ \mathsf{ifz}\,x\,?\,z\,{:}\,\mathit{nz} \ \mid\ \mathsf{noop}\to n \ \mid\ \mathsf{halt}\\
\text{programs}  & P      & =   & (\mathsf{entry}{\in}\mathsf{Node},\ \mathsf{cmd}:\mathsf{Node}\rightharpoonup\mathsf{Cmd},\ \mathsf{obs}:\mathsf{List}\,\mathsf{Var})\\
\text{stores}    & \sigma & :   & \mathsf{Var}\to\mathsf{Val}\\
\text{configs}   & c      & ::= & \langle n,\sigma\rangle\\
\end{array}$

\smallskip
\noindent$
\begin{array}{@{}l@{\ }c@{\ }l@{}}
\mathsf{succ}(n) &=& \mathbf{match}\ \mathsf{cmd}(n)\ \mathbf{with}\
  \begin{array}[t]{@{}l@{}}
  \mid (x{:=}\_)\to m \Rightarrow \{m\}\ \mid \mathsf{noop}\to m \Rightarrow \{m\}\\
  \mid \mathsf{ifz}\,x\,?\,z\,{:}\,\mathit{nz} \Rightarrow \{z,\mathit{nz}\}\ \mid \mathsf{halt} \Rightarrow \varnothing
  \end{array}\\[3pt]
\mathsf{vars}(a) &=& \mathbf{match}\ a\ \mathbf{with}\ \mid \mathsf{var}\,y \Rightarrow \{y\}\ \mid \mathsf{imm}\,v \Rightarrow \varnothing\\[3pt]
\mathsf{uses}(e) &=& \mathbf{match}\ e\ \mathbf{with}\ \mid a \Rightarrow \mathsf{vars}(a)\ \mid \odot\,a \Rightarrow \mathsf{vars}(a)\ \mid a \oplus b \Rightarrow \mathsf{vars}(a) \cup \mathsf{vars}(b)
\end{array}$
\caption{The base three address code language. A program is a partial map $\mathsf{cmd}$ from nodes to commands
along with an entry node and list of observable variables. Each command (except $\mathsf{halt}$) explicitly
identifies its successor command(s).}
\label{fig:ir-syntax}
\end{figure}

% ===== Fig. 0b — Base language expression evaluation =====
\begin{figure}[thbp]
\small
\noindent$
\begin{array}{@{}r@{\ }c@{\ }l@{}}
\mathsf{evalAtom}\,\sigma\,(\mathsf{var}\,x) & = & \sigma\,x\\
\mathsf{evalAtom}\,\sigma\,(\mathsf{imm}\,v) & = & v\\
\mathsf{eval}\,\sigma\,a & = & \mathsf{some}\,(\mathsf{evalAtom}\,\sigma\,a)\\
\mathsf{eval}\,\sigma\,(\odot a) & = & \mathsf{some}\,(\odot\,(\mathsf{evalAtom}\,\sigma\,a))\\
\mathsf{eval}\,\sigma\,(a\oplus b) & = & (\mathsf{evalAtom}\,\sigma\,a) \oplus (\mathsf{evalAtom}\,\sigma\,b)\\
\end{array}$
\caption{Expression evaluation for the base language in Figure~\ref{fig:ir-syntax}. Faults (divide or mod by zero)
are modeled as $\mathsf{none}$.}
\label{fig:ir-eval}
\end{figure}

% ===== Fig. 0c — Base language small-step semantics =====
\begin{figure}[thbp]
\small
\begin{mathpar}
\inferrule*[left=\textsc{Assign}]
  {\mathsf{cmd}(n) = (x := e)\to n' \\ \mathsf{eval}\,\sigma\,e = \mathsf{some}\,v}
  {\langle n,\sigma\rangle \to \langle n',\ \sigma[x\mapsto v]\rangle}
\and
\inferrule*[left=\textsc{Noop}]
  {\mathsf{cmd}(n) = \mathsf{noop}\to n'}
  {\langle n,\sigma\rangle \to \langle n',\sigma\rangle}
\\
\inferrule*[left=\textsc{Ifz-T}]
  {\mathsf{cmd}(n) = \mathsf{ifz}\,x\,?\,z\,{:}\,\mathit{nz} \\ \sigma\,x = 0}
  {\langle n,\sigma\rangle \to \langle z,\sigma\rangle}
\and
\inferrule*[left=\textsc{Ifz-F}]
  {\mathsf{cmd}(n) = \mathsf{ifz}\,x\,?\,z\,{:}\,\mathit{nz} \\ \sigma\,x \neq 0}
  {\langle n,\sigma\rangle \to \langle \mathit{nz},\sigma\rangle}
\end{mathpar}

\medskip
\noindent$\mathsf{WellFormed}(P) \ \Leftrightarrow\
\mathsf{entry}\in\mathrm{dom}(\mathsf{cmd})\ \ \wedge\ \
\forall\,n\,\iota,\ \mathsf{cmd}(n){=}\mathsf{some}\,\iota \Rightarrow \mathsf{succ}(n)\subseteq\mathrm{dom}(\mathsf{cmd})$

\medskip
\noindent$\mathsf{Final}\,\langle n,\sigma\rangle \Leftrightarrow \mathsf{cmd}(n)=\mathsf{halt}$\quad
$\mathsf{Faulting}\,\langle n,\sigma\rangle \Leftrightarrow \mathsf{cmd}(n)=(x{:=}e)\to n' \wedge \mathsf{eval}\,\sigma\,e=\mathsf{none}$.
\caption{Small step operational semantics $P \vdash c \to c'$ for the base language. 
$\mathsf{cmd}(n)$ matches only the operands it
needs, often eliding successors. Outcomes are only $\mathsf{halt}$, faults, or diverges.}
\label{fig:ir-semantics}
\end{figure}

\section{Base Language} 
\label{sec:base-language}

We present and implement prophecy and history variables using the 
base language in Figure~\ref{fig:ir-syntax}. The base language supports
scalar integer values and flow of control with a conditional 
branch if zero command. It represents the program in standard 
three address code representation. Each command occurs at a node $n,n'\in \mathbb{N}$
with explicit control flow --- each command identifies the next node to execute 
(see $\mathsf{succ}(n)$). Programs have an
entry node $\mathsf{entry}{\in}\mathsf{Node}$, commands 
$\mathsf{cmd}:\mathsf{Node}\rightharpoonup\mathsf{Cmd}$, and 
a list of variables $\mathsf{obs}:\mathsf{List}\,\mathsf{Var}$ whose
values are observed when the program halts at a $\mathsf{halt}$ command. 
Figure~\ref{fig:ir-eval} presents expression evaluation. 

A transition relation $\langle n,\sigma\rangle \to \langle n',\sigma'\rangle$
over configurations $c = \langle n,\sigma\rangle$ defines the base 
operational semantics (Figure~\ref{fig:ir-semantics}).
A well formed program $P$ ($\mathsf{WellFormed}(P)$ in Figure~\ref{fig:ir-semantics}) has 
no out of bounds nodes $n$. Well formed programs never get stuck;
the only outcomes are halt, diverge, and fault (the only faults are divide or mod by zero).

The base language comes with a verified compiler implementation in Lean. 
Given a program in the base language (represented in Lean data structures), the
compiler produces an assembly language representation (again in Lean
data structures) with the compilation verified to preserve the semantics
of the base language program relative to a Lean machine model of the
ARM64 ISA. There is also a verified implementation of a 
high level language with structured control flow and nested expressions 
that lowers to the base language. 
Unverified text interfaces (parser and assembly printer) complete
the compiler implementation. 
The current compiler executes the following components, all fully verified except the
parser and assembly printer:
\begin{itemize}
\item {\bf Parser:} Translates the text high level language representation into Lean data structures.
\item {\bf Lower:} Lowers the high level language Lean data structures into three address code representation. 
\item {\bf Peephole:} Performs a range of peephole transformations including simplifying operations with
constant operands and applying algebraic identities. 
\item {\bf Normalize:} Normalizes the three address code representation. Transforms the program as
necessary to enforce the following properties: 1) every successor is in range and references a command 
in the program, 2) no assignment reads the variable it defines (for example, $x := x + y \Rightarrow t := x; x := t + y$), 3) no duplicate successors ($z \neq \mathit{nz}$ in every $\mathsf{ifz}\,x\,?\,z\,{:}\,\mathit{nz}$), 
4) no branches to the entry node, 5) every node is reachable from the entry node, and 6) the entry
node is a $\mathsf{noop}$. 
\item {\bf Iterated Constant Propagation:} Runs the following sequence of transformations to a fixed point:
\begin{itemize}
\item {\bf Constant Fold:} Analyzes the program to find variables whose values are constant in 
all executions, then replaces the variable with the constant. 
\item {\bf Branch Fold:} Replaces conditional jumps ($\mathsf{ifz}\,x\,?\,z\,{:}\,\mathit{nz}$) with 
an unconditional jump to the target when $x$ is a compile time constant. The transform itself replaces
the $\mathsf{ifz}$ instruction with a $\mathsf{noop}$ whose successor is the branch target. 
\item {\bf Unreachable Code Elimination:} Removes any instructions that are not reachable from the entry node. 
\end{itemize}
\item {\bf Normalize:} Normalizes the three address code representation after iterated constant propagation. 
\item {\bf Lazy Code Motion:} Performs lazy code motion.
\item {\bf Partial Dead Code Elimination:} Performs partial dead code elimination.
\item {\bf Cleanup:} Removes superfluous $\mathsf{noop}$ instructions. 
\item {\bf Code Generation:} Lowers optimized three address code program to an assembly program represented
as Lean data structures. 
\item {\bf Assembly Printer:} Generates a text assembly language representation of the program. 
\end{itemize}

\section{Partial Dead Code Elimination (PDCE)}
\label{sec:pdce}

Conceptually, Partial Dead Code Elimination (PDCE) moves assignments forward
with the flow of control to be closer to their uses~\cite{DBLP:conf/pldi/KnoopRS94}. Moving assignments
through control flow branches can enable the elimination of assignments
that are dead on one branch but not the other --- assignments to nonobservable
variables $x \notin \mathsf{obs}$ that move all the
way to $\mathsf{halt}$ statements can be dropped from the program. 

% GENERATED by script/gsl2tex.py — do not edit by hand; regenerate via script/gen-gsl-figures.sh
% Requires \usepackage{listings}; reads the verbatim block from figures/pdce-spec.gsltex.
\ifdefined\gslListingsReady\else
  \lstdefinelanguage{GSL}{morekeywords={analysis,prophecy,history,predict,check,update,seed,within,when,meets},morecomment=[l]{--},sensitive=true}
  \lstdefinestyle{gslstyle}{language=GSL,basicstyle=\sffamily\small,keywordstyle=\bfseries,
    columns=fullflexible,keepspaces=true,extendedchars=true,showstringspaces=false,
    breaklines=true,breakatwhitespace=true,breakindent=6em,
    literate=
    {π}{{$\pi$}}1
    {η}{{$\eta$}}1
    {τ}{{$\tau$}}1
    {ₐ}{{${}_{a}$}}1
    {ₚ}{{${}_{p}$}}1
    {ᵤ}{{${}_{u}$}}1
    {∀}{{$\forall$}}1
    {→}{{$\to$}}1
    {∅}{{$\emptyset$}}1
    {∖}{{$\setminus$}}1
    {∩}{{$\cap$}}1
    {∪}{{$\cup$}}1
    {⊆}{{$\subseteq$}}1
    {∈}{{$\in$}}1
    {·}{{$\cdot$}}1
    {⇒}{{$\Rightarrow$}}1}
  \let\gslListingsReady\relax
\fi
\begin{figure}[thbp]
\begin{lstlisting}[style=gslstyle]
analysis PDCE {
  prophecy Live π : Variables[Var] {
    predict : ∀ n→n'. π(n') ⊆ π(n) ∪ defVars(n)
    check   : ∀ n→n'. rhsVars(n) ⊆ π(n)  when  defVars(n) meets π(n')
    check   : ∀ n. condVars(n) ⊆ π(n)
    seed    : liveSeed ⊆ π(final)
    within  : ∀ n. π(n) ⊆ allVars
  }
  history Sink η : Assignments[Asgn] {
    update : ∀ n→n'. η(n') ⊆ born(n) ∪ (η(n) ∩ pass(n))
    seed   : η(entry) ⊆ sinkSeed
    within : ∀ n. η(n) ⊆ allAsgns
  }
}
\end{lstlisting}
\caption{PDCE analysis specification.}
\label{fig:pdce-spec}
\end{figure}

% ===== Fig. 0d — PDCE node local sets (from Pdce.gsl / PdceDefs.lean) =====
\begin{figure}[thbp]
\small
\noindent$
\begin{array}{@{}l@{\ }c@{\ }l@{}}
\mathsf{defVars}(n)  & = & \mathbf{match}\ \mathsf{cmd}(n)\ \mathbf{with}\ \mid x{:=}e \Rightarrow \{x\}\ \mid \_ \Rightarrow \varnothing\\
\mathsf{rhsVars}(n)  & = & \mathbf{match}\ \mathsf{cmd}(n)\ \mathbf{with}\ \mid x{:=}e \Rightarrow \mathsf{uses}(e)\ \mid \_ \Rightarrow \varnothing\\
\mathsf{condVars}(n) & = & \mathbf{match}\ \mathsf{cmd}(n)\ \mathbf{with}\ \mid \mathsf{ifz}\,x \Rightarrow \{x\}\ \mid \_ \Rightarrow \varnothing\\
\mathsf{born}(n)     & = & \mathbf{match}\ \mathsf{cmd}(n)\ \mathbf{with}\ \mid x{:=}e \Rightarrow \{\langle x,e\rangle\}\ \mid \_ \Rightarrow \varnothing\\
\mathsf{allAsgns}    & \stackrel{\triangle}{=} & \textstyle\bigcup_{n} \mathsf{born}(n)\quad\text{(the assignments computed in $P$)}\\
\mathsf{kills}(n,\langle w,f\rangle) & = & w \in \mathsf{rhsVars}(n) \cup \mathsf{condVars}(n)\ \vee\ w \in \mathsf{defVars}(n)\ \vee\ \mathsf{defVars}(n) \cap \mathsf{uses}(f) \neq \varnothing\\
\mathsf{pass}(n)     & = & \{\, a \in \mathsf{allAsgns} \mid \neg\,\mathsf{kills}(n,a) \,\}\\
\mathsf{allVars}     & \stackrel{\triangle}{=} & \mathsf{obs} \cup \textstyle\bigcup_{n} (\mathsf{rhsVars}(n) \cup \mathsf{condVars}(n) \cup \mathsf{defVars}(n))\\
\mathsf{liveSeed}    & = & \mathsf{obs}\\
\mathsf{sinkSeed}    & = & \varnothing
\end{array}$
\caption{PDCE node local sets and constants.}
\label{fig:pdce-locals}
\end{figure}

\subsection{PDCE Analysis}

We formulate the PDCE analysis with the combination of a prophecy variable
$\pi$ and history variable $\eta$. Figure~\ref{fig:pdce-spec} specifies the 
PDCE analysis, specifying the prophecy and history variables as subset inclusion constraints over 
the nodes and transitions of the base operational semantics (definitions of node local sets
and constants in Figure~\ref{fig:pdce-locals}). At the start of each node $n$, the 
prophecy variable $\pi(n)$ predicts the variables that are live at $n$.
The history variable $\eta(n)$ records the assignments that have currently
moved to $n$ and are eligible to move further. 
The $\mathsf{predict}$ clause predicts which variables $\pi(n')$ will be
live after $n$ executes. Specifically, in the predict clause
\begin{equation}\label{eq:pdce-predict}
  \mathsf{predict}\; : \;  \langle n,\_\rangle \to \langle n',\_\rangle\ \Rightarrow\ \pi(n') \subseteq \pi(n) \cup \mathsf{defVars}(n)
\end{equation}
the prediction $\pi(n') \subseteq \pi(n) \cup \mathsf{defVars}(n)$
predicts that, after $n$ executes, any variables that $n$ defines (i.e., $\mathsf{defVars}(n)$) may become live 
and some of the incoming live variables $\pi(n)$ may no
longer be live (when $n$ contains the last read of the variable). 
For example, if $\mathsf{cmd}(n) = x:=y \oplus z$, $x$ may become live after $n$
and $y$ and $z$ may no longer be live after $n$.
The check clause
\begin{equation}
\begin{array}{@{}l@{\;:\;}l@{}}
\mathsf{check} & \langle n,\_\rangle \to \langle n',\_\rangle \;\Rightarrow\; \mathsf{rhsVars}(n) \subseteq \pi(n)\ \ \mathbf{when}\ \ \mathsf{defVars}(n) \cap \pi(n') \neq \varnothing\\
\end{array}
\end{equation}
requires that, if $x$ is live after an assignment $x{:=}e$, (i.e., $\{ x \} \cap \pi(n') \neq \varnothing$) then the variables in $e$ (i.e., 
$\mathsf{rhsVars}(n)$) must be live before the assignment $x{:=}e$ (because they are used to 
compute the value of $x$). This check enables the elimination of assignments
that are {\em faint} --- transitively dead because they are used 
only by other dead code. The check clause 
\begin{equation}
\begin{array}{@{}l@{\;:\;}l@{}}
\mathsf{check} & \mathsf{condVars}(n) \subseteq \pi(n)
\end{array}
\end{equation}
requires that the variable $x$ read by a conditional branch must be live at the conditional branch. 

The update clause updates the history variable $\eta(n)$, which carries moving
assignments $\langle w,f\rangle \in \eta(n)$. 
\begin{equation}\label{eq:pdce-update}
\mathsf{update} \; : \; \langle n,\_\rangle \to \langle n',\_\rangle\ \Rightarrow\ \eta(n') \subseteq \mathsf{born}(n) \cup (\eta(n) \cap \mathsf{pass}(n))
\end{equation}

At an assignment statement $\mathsf{cmd}(n) = x:=e$, the assignment $\langle x,e\rangle$ is born (i.e., $\mathsf{born}(n) = \{ \langle x,e\rangle\}$),
so the update enables the assignment to move forward via $\eta(n') \subseteq \mathsf{born}(n)$. The update also 
prevents a moving assignment $\langle w,f\rangle \in \eta(n)$ from passing through 
$x{:=}e$ when $\langle w,f\rangle \notin \mathsf{pass}(n)$, which simplifies to 
either $w \in \mathsf{uses}(e)$ (i.e., the assignment
uses $w$), $w = x$ (i.e., the assignment redefines $w$ so $f$ is no longer
the right value for $w$ after the assignment), or $x \in \mathsf{uses}(f)$ (i.e., 
one of the operands in $f$ is redefined so that $w$ no longer holds the right
value of $f$ after the assignment). 

% ===== Fig. 2 — PDCE sinking placement (insert / delete) =====
\begin{figure}[thbp]
\small
\begin{mathpar}
\inferrule*[left=\textsc{Ins-entry}]
  {\langle w,f\rangle \in \eta(n) \\ \langle w,f\rangle \in \mathsf{blocked}(n) \\ w \in \pi(n)}
  {w := f\ \in\ \mathsf{entry}(n)}
\\
\inferrule*[left=\textsc{Ins-edge}]
  {n' \in \mathsf{succ}(n) \\ \langle w,f\rangle \in \mathsf{delayedExit}(n) \\ \langle w,f\rangle \notin \eta(n') \\ w \in \pi(n')}
  {w := f\ \in\ \mathsf{edge}(n,n')}
\\
\inferrule*[left=\textsc{Delete}]
  {\mathsf{cmd}(n) = (x := e)}
  {\mathsf{ctrl}(n) =\ \mathsf{noop}}
\end{mathpar}

\medskip
\noindent$
\begin{array}{@{}l@{\ }c@{\ }l@{}}
\mathsf{blocked}(n)     & = & \mathbf{match}\ \mathsf{cmd}(n)\ \mathbf{with}\ \mid \mathsf{halt} \Rightarrow \mathsf{allAsgns}\ \mid \_ \Rightarrow \neg\,\mathsf{pass}(n)\\
\mathsf{delayedExit}(n) & = & \mathsf{born}(n) \cup (\eta(n) \cap \mathsf{pass}(n))
\end{array}$
\caption{PDCE assignments $w := f$ inserted before nodes ($\textsc{Ins-entry}$) and on edges ($\textsc{Ins-edge}$).}
\label{fig:pdce-insert}
\end{figure}

% ===== Fig. X — PDCE worked example: node insert + edge insert + elimination =====
% Requires in the main preamble: \usepackage{tikz}\usetikzlibrary{arrows.meta,calc}
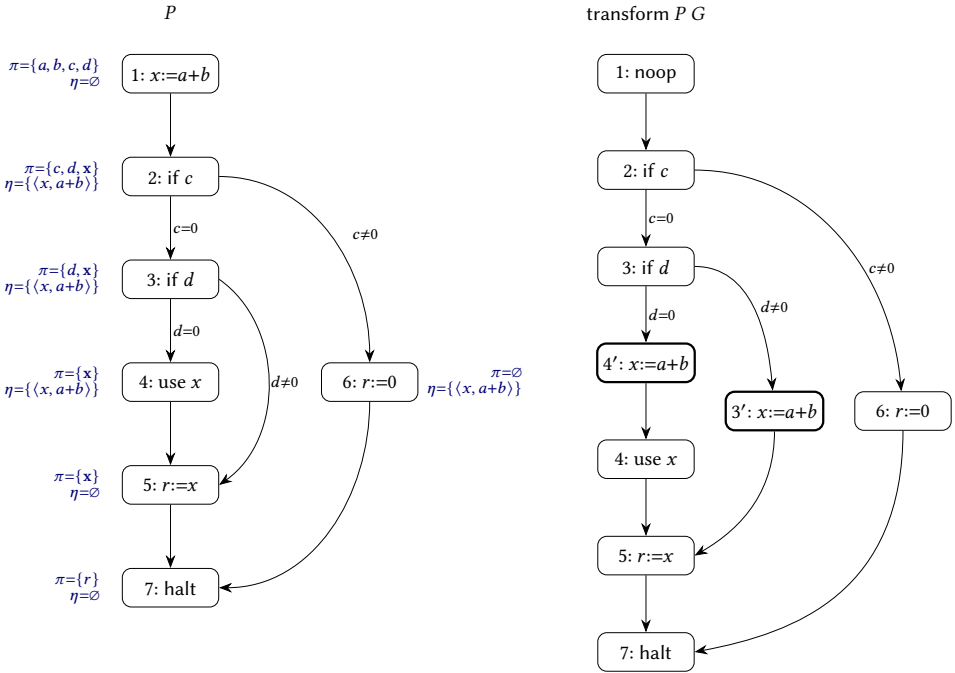
\begin{figure*}[thbp]
\centering
\small
\scalebox{0.85}{%
\begin{tikzpicture}[
  cfg/.style   ={draw, rounded corners, font=\small, inner sep=2.5pt, minimum width=15mm, minimum height=6mm, align=center, fill=white},
  ins/.style   ={cfg, line width=1pt},
  ed/.style    ={-{Stealth[length=2mm]}, font=\scriptsize, inner sep=1pt},
  ann/.style   ={font=\scriptsize, color=blue!45!black, align=right, inner sep=1pt},
  eins/.style  ={draw=green!55!black, very thick, fill=green!12, rounded corners, font=\scriptsize, inner sep=2.5pt},
  lbl/.style   ={font=\scriptsize\itshape, color=green!45!black, inner sep=1pt, align=center},
]

% ---------------- ORIGINAL ----------------
\node[cfg] (o1) at (0,0)      {$1{:}\ x{:=}a{+}b$};
\node[cfg] (o2) at (0,-1.6)   {$2{:}\ \mathsf{if}\ c$};
\node[cfg] (o3) at (0,-3.2)   {$3{:}\ \mathsf{if}\ d$};
\node[cfg] (o4) at (0,-4.8)   {$4{:}\ \text{use}\ x$};
\node[cfg] (o5) at (0,-6.4)   {$5{:}\ r{:=}x$};
\node[cfg] (o7) at (0,-8.0)   {$7{:}\ \mathsf{halt}$};
\node[cfg] (o6) at (3.1,-4.8) {$6{:}\ r{:=}0$};

\draw[ed] (o1) -- (o2);
\draw[ed] (o2) -- (o3) node[midway,right]{$c{=}0$};
\draw[ed] (o3) -- (o4) node[midway,right]{$d{=}0$};
\draw[ed] (o4) -- (o5);
\draw[ed] (o5) -- (o7);
\draw[ed] (o3.east) to[bend left=55] node[midway,right]{$d{\neq}0$} (o5.east);
\draw[ed] (o2.east) to[out=0,in=90] node[pos=0.5,right=9pt]{$c{\neq}0$} (o6.north);
\draw[ed] (o6.south) to[out=-90,in=0] (o7.east);

% analysis annotations (left of the spine)
\node[ann,anchor=east] at (-1.05,0)    {$\pi{=}\{a,b,c,d\}$\\[-1pt]$\eta{=}\varnothing$};
\node[ann,anchor=east] at (-1.05,-1.6) {$\pi{=}\{c,d,\mathbf{x}\}$\\[-1pt]$\eta{=}\{\langle x, a{+}b\rangle\}$};
\node[ann,anchor=east] at (-1.05,-3.2) {$\pi{=}\{d,\mathbf{x}\}$\\[-1pt]$\eta{=}\{\langle x, a{+}b\rangle\}$};
\node[ann,anchor=east] at (-1.05,-4.8) {$\pi{=}\{\mathbf{x}\}$\\[-1pt]$\eta{=}\{\langle x, a{+}b\rangle\}$};
\node[ann,anchor=east] at (-1.05,-6.4) {$\pi{=}\{\mathbf{x}\}$\\[-1pt]$\eta{=}\varnothing$};
\node[ann,anchor=east] at (-1.05,-8.0) {$\pi{=}\{r\}$\\[-1pt]$\eta{=}\varnothing$};
\node[ann,anchor=west] at (3.95,-4.8)  {$\pi{=}\varnothing$\\[-1pt]$\eta{=}\{\langle x, a{+}b\rangle\}$};

\node[font=\bfseries] at (0,0.95) {$P$};

% ---------------- TRANSFORMED ----------------
\begin{scope}[xshift=7.4cm]
\node[cfg]  (t1)  at (0,0)      {$1{:}\ \mathsf{noop}$};
\node[cfg]  (t2)  at (0,-1.5)   {$2{:}\ \mathsf{if}\ c$};
\node[cfg]  (t3)  at (0,-3.0)   {$3{:}\ \mathsf{if}\ d$};
\node[ins]  (t4p) at (0,-4.5)    {$4'{:}\ x{:=}a{+}b$};
% edge-insert 3' centred between 3 and 5; node 6 centred between 2 and 7 (both y=-5.25):
\node[ins]  (t3p) at (2.0,-5.25) {$3'{:}\ x{:=}a{+}b$};
\node[cfg]  (t6)  at (4.0,-5.25) {$6{:}\ r{:=}0$};
\node[cfg]  (t4)  at (0,-6.0)    {$4{:}\ \text{use}\ x$};
\node[cfg]  (t5)  at (0,-7.5)    {$5{:}\ r{:=}x$};
\node[cfg]  (t7)  at (0,-9.0)    {$7{:}\ \mathsf{halt}$};

\draw[ed] (t1) -- (t2);
\draw[ed] (t2) -- (t3) node[midway,right]{$c{=}0$};
\draw[ed] (t3) -- (t4p) node[midway,right]{$d{=}0$};
\draw[ed] (t4p) -- (t4);
\draw[ed] (t4) -- (t5);
\draw[ed] (t5) -- (t7);
\draw[ed] (t3.east) to[out=0,in=95] node[pos=0.5,right=2pt]{$d{\neq}0$} (t3p);
\draw[ed] (t3p.south) to[out=-90,in=30] (t5.east);
\draw[ed] (t2.east) to[out=0,in=95] node[pos=0.62,right=2pt]{$c{\neq}0$} (t6);
\draw[ed] (t6.south) .. controls +(0,-2.5) and +(1.6,0.3) .. (t7.east);

\node[font=\bfseries] at (0,0.95) {$\mathsf{transform}\ P\ G$};
\end{scope}

\end{tikzpicture}%
}

\caption{PDCE example transform. The assignment $x := a+b$ is moved from node $1$ and inserted
before node $4$ (new node $4'$ in the transformed program) and on edge $3$ to $5$ (new node
$3'$ in the transformed program). The transform eliminates $x := a+b$ on the path $1,2,6,7$.}
\label{fig:pdce-example}
\end{figure*}
% ===== Fig. — PDCE step-by-step forward simulation / refinement table =====
% Companion to fig:pdce-example (same program, nodes, and inserts 4', 3').
% The refinement mapping is LIVENESS-GUARDED (matches the mechanized `Match`, PDCE.md 7.2,
% BaseLanguage/PDCE/Correctness.lean): a variable is constrained only where it is live ---
%   live & in-flight    : some(sigma_o x) = eval sigma_t e
%   live & not-in-flight : sigma_o x = sigma_t x
%   dead : no constraint.
% On the eliminated path 2,6,7 the moved x is dead (node 6 has pi = {}, at halt only r is live),
% so the mapping makes NO claim on x there; the observable r agrees.
\begin{figure*}[thbp]
\centering
\footnotesize\setlength{\tabcolsep}{5pt}\renewcommand{\arraystretch}{1.6}
\begin{tabular}{llcc}
\hline
 & & \multicolumn{2}{c}{Refinement mapping} \\
\cline{3-4}
\noalign{\vskip 3pt}
\shortstack[l]{Original\\ program step} & \shortstack[l]{Transformed\\ program steps} & \shortstack{At entry\\ configuration} & \shortstack{At exit\\ configuration} \\
\hline
$\langle 1,\sigma_o^1\rangle{\to}\langle 2,\sigma_o^2\rangle$ & $\langle 1,\sigma_t^1\rangle{\to}\langle 2,\sigma_t^2\rangle$ & $\sigma_o^1 x = \sigma_t^1 x$ & $\mathsf{some}\,(\sigma_o^2 x)=\mathsf{eval}\,\sigma_t^2\,(a{+}b)$ \\
$\langle 2,\sigma_o^2\rangle{\to}\langle 3,\sigma_o^3\rangle$~{\scriptsize$(c{=}0)$}    & $\langle 2,\sigma_t^2\rangle{\to}\langle 3,\sigma_t^3\rangle$ & $\mathsf{some}\,(\sigma_o^2 x)=\mathsf{eval}\,\sigma_t^2\,(a{+}b)$ & $\mathsf{some}\,(\sigma_o^3 x)=\mathsf{eval}\,\sigma_t^3\,(a{+}b)$ \\
$\langle 2,\sigma_o^2\rangle{\to}\langle 6,\sigma_o^6\rangle$~{\scriptsize$(c{\neq}0)$} & $\langle 2,\sigma_t^2\rangle{\to}\langle 6,\sigma_t^6\rangle$ & $\mathsf{some}\,(\sigma_o^2 x)=\mathsf{eval}\,\sigma_t^2\,(a{+}b)$ & $\pi(6){=}\varnothing$ \\
$\langle 3,\sigma_o^3\rangle{\to}\langle 4,\sigma_o^4\rangle$~{\scriptsize$(d{=}0)$}    & $\langle 3,\sigma_t^3\rangle{\to}\langle 4',\sigma_t^{4'}\rangle$ & $\mathsf{some}\,(\sigma_o^3 x)=\mathsf{eval}\,\sigma_t^3\,(a{+}b)$ & $\mathsf{some}\,(\sigma_o^4 x)=\mathsf{eval}\,\sigma_t^{4'}\,(a{+}b)$ \\
$\langle 3,\sigma_o^3\rangle{\to}\langle 5,\sigma_o^5\rangle$~{\scriptsize$(d{\neq}0)$} & $\langle 3,\sigma_t^3\rangle{\to}\langle 3',\sigma_t^{3'}\rangle{\to}\langle 5,\sigma_t^5\rangle$ & $\mathsf{some}\,(\sigma_o^3 x)=\mathsf{eval}\,\sigma_t^3\,(a{+}b)$ & $\sigma_o^5 x = \sigma_t^5 x$ \\
$\langle 4,\sigma_o^4\rangle{\to}\langle 5,\sigma_o^5\rangle$ & $\langle 4',\sigma_t^{4'}\rangle{\to}\langle 4,\sigma_t^4\rangle{\to}\langle 5,\sigma_t^5\rangle$ & $\mathsf{some}\,(\sigma_o^4 x)=\mathsf{eval}\,\sigma_t^{4'}\,(a{+}b)$ & $\sigma_o^5 x = \sigma_t^5 x$ \\
$\langle 5,\sigma_o^5\rangle{\to}\langle 7,\sigma_o^7\rangle$ & $\langle 5,\sigma_t^5\rangle{\to}\langle 7,\sigma_t^7\rangle$ & $\sigma_o^5 x = \sigma_t^5 x$ & $\sigma_o^7 r = \sigma_t^7 r$ \\
$\langle 6,\sigma_o^6\rangle{\to}\langle 7,\sigma_o^7\rangle$ & $\langle 6,\sigma_t^6\rangle{\to}\langle 7,\sigma_t^7\rangle$ & $\pi(6){=}\varnothing$ & $\sigma_o^7 r = \sigma_t^7 r$ \\
\hline
\end{tabular}
\caption{Example simulation relation for the PDCE transform of Figure~\ref{fig:pdce-example}, including
refinement mapping. The mapping constrains only live variables. For live $a$, $b$, $c$, and $d$, the
mapping is the identity (not shown). }
\label{fig:pdce-steps}
\end{figure*}

\subsection{PDCE Transform} 

The PDCE transform turns all assignments into $\mathsf{noop}$s, then reinserts
the assignments at optimal late insertion points either before a node $n$ or on an outgoing
edge from $n$ to $n'$. Figure~\ref{fig:pdce-insert} presents the definitions 
of the set of assignments $w := f\ \in\ \mathsf{entry}(n)$ (rule \textsc{Ins-entry}) inserted before $n$ and 
the set of assignments $w := f\ \in\ \mathsf{edge}(n,n')$ inserted on the edge from $n$ to $n'$ 
(rule \textsc{Ins-edge}).
First, the transform inserts only {\em live} assignments 
$w := f$, i.e., assignments with $w$ live at the insertion point ($w \in \pi(n)$, $w \in \pi(n')$)
--- this constraint eliminates
dead assignments. A live assignment $w := f$ is inserted before $n$ if it can move to $n$ ($\langle w,f\rangle \in \eta(n)$)
but is blocked from moving any further ($\langle w,f\rangle \in \mathsf{blocked}(n)$), 
either because $n$ uses $w$, redefines $w$, redefines a 
variable in $f$, or because $n$ is $\mathsf{halt}$ (where every assignment is blocked to ensure
that all still live assignments are inserted before the $\mathsf{halt}$).
A live assignment $w := f$ is inserted on an edge from $n$ to $n'$ if it
can move to the program point after $n$ ($\langle w,f\rangle \in \mathsf{delayedExit}(n)$) but 
not to $n'$ ($\langle w,f\rangle \notin \eta(n')$).

The \textsc{Delete} rule changes all of the assignments in the original program into 
$\mathsf{noop}$s (the notation $\mathsf{ctrl}(n) =\ \mathsf{noop}$ indicates that the
command at $n$ is set to $\mathsf{noop}$) and the transformation inserts the 
assignments as indicated by the $\mathsf{entry}(n)$ and $\mathsf{edge}(n,n')$ sets.
A verified cleanup phase that executes after PDCE removes $\mathsf{noop}$s.

Figure~\ref{fig:pdce-example} presents an example PDCE transform. The assignment
$x := a+b$ at node $1$ in the original program $P$ is moved downward to just before 
node $4$ (at node $4'$ in the transformed program) and inserted along the edge 
from $3$ to $5$ (at node $3'$ in the transformed program). The transform 
shrinks the live ranges and eliminates the evaluation of $x := a+b$ on the path $1,2,6,7$. 

The PDCE correctness and optimality proofs work with the concept of the {\em image} of the original
program in the transformed program. Recall that the transform inserts
a (possibly empty) block of assignments before each node $n$. That block, plus $n$, is 
the {\em image} of $n$ in the transformed program. The {\em entry} of the image of $n$ is the
node of the first assignment in the inserted block ($n$ if the block is empty). 
The transform also inserts a (possibly empty)
block of assignments on each edge $n \rightarrow n'$. The image of $n$, the edge from $n$ into 
the block, the block of assignments, plus the edge from the last assignment in the block to the entry 
of the image of $n'$, is the image of the edge $n \rightarrow n'$. We denote the
entry of the image of $n$ in the transformed program $T$ as $\mathsf{block}(n,T)$
and $\mathsf{block}(n)$ when $T$ is obvious from context. 

\subsection{PDCE Correctness}

The PDCE correctness theorem states that if the original program halts, then
the transformed program also halts and the values of the observable variables
are the same in both executions. There is also a theorem that the transform preserves divergences.
The transform does not necessarily preserve faults --- the transform may push a faulting 
command (divide or mod by zero) past a conditional branch and out of the execution.

% transform_preserves_halt
% transform_preserves_diverges

The PDCE correctness proof is a forward simulation over the original program $P$
and transformed program $T$. The proof matches each step 
$\langle n_o,\sigma_o\rangle \to \langle n_o',\sigma_o'\rangle$ from the original 
program against potentially multiple steps 
$\langle n_t,\sigma_t\rangle \to \cdots \to \langle n_t',\sigma_t'\rangle$ 
from the transformed program (the transform 
may have moved multiple assignments to the program points covered by the 
multiple steps in the transformed program). Here $n_t \to \cdots \to n_t'$ is the 
image of $n_o \to n_o'$, so that $n_t = \mathsf{block}(n_o)$ and
$n_t' = \mathsf{block}(n_o')$.  

The proof maintains the invariant that at corresponding
configurations during the simulation ($\langle n_o,\sigma_o\rangle$ corresponds to $\langle n_t,\sigma_t\rangle$, 
$\langle n_o',\sigma_o'\rangle$ corresponds to $\langle n_t',\sigma_t'\rangle$), the 
following refinement mapping holds for each variable $x$ at corresponding configurations
$\langle n_o,\sigma_o\rangle$, $\langle n_t,\sigma_t\rangle$:
\[
\begin{array}{l@{\qquad}l}
    \mathsf{some}\,(\sigma_o\,x) \;=\; \mathsf{eval}\,\sigma_t\,e
      & \text{if } x \in \pi(n_o) \text{ and } \langle x, e \rangle \in \eta(n_o),\\[4pt]
    \sigma_o\,x \;=\; \sigma_t\,x
      & \text{if } x \in \pi(n_o) \text{ and } \forall e.\ \langle x, e \rangle \notin \eta(n_o),\\[4pt]
    \text{(unconstrained)}
      & \text{if } x \notin \pi(n_o).
\end{array}
\]
i.e., the corresponding stores $\sigma_o$ and $\sigma_t$ agree on the value of every \emph{live}
variable $x$ that has no assignment in flight at $n_o$. If a live $x$ has an assignment
$\langle x, e \rangle$ in flight, then $e$ in $\sigma_t$ evaluates to the same value as $x$ in
$\sigma_o$; a \emph{dead} variable ($x \notin \pi(n_o)$) is left unconstrained --- it may legitimately
differ between $\sigma_o$ and $\sigma_t$ (e.g., a partially dead computation eliminated on the current path).

The invariant also maintains that the assignments in $\eta(n_o)$ are \emph{noninterfering} --- for any two
distinct assignments $\langle w, f\rangle, \langle w', f'\rangle \in \eta(n_o)$,
$w \neq w'$, $w \notin \mathsf{vars}(f')$, and $w' \notin \mathsf{vars}(f)$. With this constraint
the assignments in each block can be reordered (which complicates the statement of the optimality
theorems in Section~\ref{subsec:pdce-opt}).

A case analysis on the instruction at $n_o$ in the original program
uses the fact that the prophecy variable $\pi(n_o)$ and history variable $\eta(n_o)$ satisfy
the prophecy and history variable conditions in the augmented operational semantics
to push the invariant through the step (in the original program) and its corresponding
image (in the transformed program). Note that the correctness proof
only requires a valid analysis result $G$, not an extremal
analysis result. This proof structure highlights the automated reasoning benefits that 
the tight coupling of the prophecy
and history variables to the operational semantics can deliver. 

Figure~\ref{fig:pdce-steps} presents original program steps and transformed program
sequence of steps, along with a refinement mapping, that illustrate a simulation relation
between the original and transformed programs in Figure~\ref{fig:pdce-example}.

\subsection{PDCE Optimality}
\label{subsec:pdce-opt}

The PDCE optimality theorems compare executions of two transformed
programs $T$ and $T'$ --- $T = \mathsf{transform}\; P\; G$ (the original program $P$
transformed with the extremal analysis result $G$) and $T' = \mathsf{transform}\; P\; G'$
(the original program transformed with an arbitrary valid analysis result $G'$). 
% All work with the concept of a {\em skeleton} --- the image of the original program in the transformed program. Recall that the transform conceptually preserves the nodes of the original program (potentially converted to $\mathsf{noop}$s) and inserts blocks of moved assignments before nodes and on edges of the original program.  We call the image of the nodes of the original program in the transformed program the {\em skeleton} nodes. 

The headline optimality theorems are stated over execution traces of the original program $P$ and use the 
ghost variables $\pi$ and $\eta$ to reason about the 
corresponding traces of $T$ and $T'$ constructed by inserting 
blocks of moved assignments into the original program traces.
Because, in general, assignments in the same block can be reordered, the theorems
are proved to hold only at entries of the images of nodes in the transformed program:
\begin{itemize}
\item {\bf Assignment Optimality:} For any assignment $a$, all traces of $T'$ execute $a$ at
least as many times as the corresponding trace of $T$.
% matCount_le_any - one trace of original program, reason about T and T' with ghost variables. 
% transform_execCountOrig_le_any - builds traces of two transformed programs. 
\item {\bf Live Range Optimality:} For any variable $x$ and any trace of $T'$, the live
ranges of $x$ in $T'$ as specified by $G'$ are at least as large as the live ranges of $x$ in the corresponding
trace of $T$ as specified by $G$. Note the specified by $G'$ and $G$ qualification --- the live
ranges are computed with respect to the uses of $x$ in the original program. While these uses appear
as nodes in $T$ and $T'$, they may be converted to $\mathsf{noop}$s with the uses moved.
These moves can increase the live ranges of $x$ in $T$ and $T'$ so that the live ranges in the 
transformed program $T'$ may be smaller than the live ranges in $T$. If the uses of $x$ do not move ---
for example, if uses are source-level statements and only compiler introduced temporaries move --- 
then the live ranges in $T'$ are at least as large as the live ranges in $T$.

% --- and we have a proof of this theorem. 
% pathVarLiveRange_le: one original-program trace; reasons about T and T' via the ghosts G, G'.
% transform_regOccOrig_le_any: runs the two transformed programs T, T' (each folds to pathVarLiveRange).
% transform_honestLiveRange_le: if reads of x don't move, live range in T <= live range in T', path in original program
% transform_regOccExtOrig_le_any - if reads of x don't move, live range in T <= live range in T', paths in transformed program
% 

\end{itemize}
We have also stated versions of these theorems directly over explicit traces 
of the transformed programs and proved these versions as well.
% The proofs here conceptually use a triple simulation --- a fuel-based simulation constructs a trace of the original program, then translates the trace into the transformed then use the skeletons in the transformed programs to construct corresponding transformed program traces and prove the theorems over those traces. 

All of these theorems rely heavily on trace based arguments that leverage 
prophecy and history variable properties as defined in the 
transition rules of the augmented operational semantics. The proofs of live range optimality
traverse an arbitrary trace of the original program, using the prophecy and
history variables to reason about the live ranges of variables in the transformed programs $T$ and $T'$.
A node $n$ is in the live range of a variable $x$ if 1) $x$ is live at $n$ (i.e., $x \in \pi(n)$) and 2) 
$x$ is not in flight at $n$ (i.e., $\not \exists \langle x, e \rangle \in \eta(n)$) --- in other words,
$x$ will be read at or downstream of $n$ and the assignment whose value will be read has already been 
inserted upstream of $n$.  A key lemma states that if $\mathsf{block}(n,T)$ (here
$T$ is the program transformed with extremal $G$) is in the live range of $x$,
then $\mathsf{block}(n,T')$ (here $T'$ is the program transformed with valid but not necessarily extremal $G'$)
is also in the live range of $x$. This lemma follows immediately from the fact that 
for all $n$, $\pi(n) \subseteq \pi'(n)$ and $\eta'(n) \subseteq \eta(n)$ for extremal $\pi, \eta$ and
arbitrary valid $\pi', \eta'$.

The assignment optimality proofs reason about the values of the prophecy and history variables to count
the number of times an assignment $a$ is executed in the two transformed programs $T$ and $T'$
for corresponding arbitrary traces starting at the entry of the program. A key aspect of the proof
is that $a$ may be inserted later in the trace for $T$ than for $T'$ but never earlier. The proof
is an induction over the trace and uses the history and prophecy variables at each step to compare
insertion points for $a$ in the two programs, ensuring that the number of times $a$ is inserted in 
any prefix of the trace of $T$ is always at most the number of times $a$ is inserted in 
corresponding prefix of the trace of $T'$.

% ===== Fig. 6 — LCM hoisting placement (insert / replace) =====
\begin{figure}[thbp]
\small
\begin{mathpar}
\inferrule*[left=\textsc{Ins-entry}]
  {e \in \mathsf{insertBefore}(n)}
  {t_e := e\ \in\ \mathsf{entry}(n)}
\\
\inferrule*[left=\textsc{Ins-edge}]
  {m \in \mathsf{succ}(n) \\ e \in \mathsf{insertEdge}(n,m)}
  {t_e := e\ \in\ \mathsf{edge}(n,m)}
\\
\inferrule*[left=\textsc{Replace}]
  {\mathsf{cmd}(n) = (x := e) \\ \mathsf{numbered}(e) \\ e \in \mathsf{recoverable}(n)}
  {\mathsf{ctrl}(n) =\ x := t_e}
\\
\inferrule*[left=\textsc{Keep}]
  {\mathsf{cmd}(n) = (x := e) \\ \neg(\mathsf{numbered}(e) \wedge e \in \mathsf{recoverable}(n))}
  {\mathsf{ctrl}(n) =\ x := e}
\end{mathpar}

\medskip
\noindent$
\begin{array}{@{}l@{\ }c@{\ }l@{}}
\mathsf{insertEdge}(n,m)  & = & \mathsf{latestEdge}(\pi_a,\eta_a,\eta_p)(n,m) \cap \tau_u(n)\\
\mathsf{insertBefore}(n)  & = & \big(\mathsf{latestNode}(\eta_p,\tau_p)(n) \cap \tau_u(n)\big) \setminus \textstyle\bigcup_{m \in \mathsf{succ}(n)} \mathsf{latestEdge}(\pi_a,\eta_a,\eta_p)(n,m)\\
\mathsf{recoverable}(n)   & = & \pi_u(n) \cup \mathsf{insertBefore}(n)
\end{array}$
\caption{LCM expression evaluations $t_e := e$ inserted before nodes ($\textsc{Ins-entry}$) and on edges ($\textsc{Ins-edge}$).}
\label{fig:lcm-insert}
\end{figure}

% ===== LCM node local sets & placement functions (from analyses/lcm/LcmDefs.lean) =====
\begin{figure*}[thbp]
\small
\noindent$
\begin{array}{@{}l@{\ }c@{\ }l@{}}
\textbf{Node local sets} & &\\[2pt]
\mathsf{ue}(n)       & = & \mathbf{match}\ \mathsf{cmd}(n)\ \mathbf{with}\ \mid x{:=}e \Rightarrow \mathsf{numbered}(e)\,?\,\{e\}:\varnothing \ \mid \_ \Rightarrow \varnothing\\
\mathsf{numbered}(e) & = & \mathbf{match}\ e\ \mathbf{with}\ \mid a \Rightarrow \mathsf{false}\ \mid \odot\,a \Rightarrow \mathsf{true}\ \mid a \oplus b \Rightarrow \mathsf{true}\\
\mathsf{allExprs}    & \stackrel{\triangle}{=} & \textstyle\bigcup_{n} \mathsf{ue}(n)\quad\text{(the expressions computed in $P$)}\\
\mathsf{pass}(n)     & = & \{\, e \in \mathsf{allExprs} \mid \mathsf{def}(n) \cap \mathsf{uses}(e) = \varnothing \,\}\\
\mathsf{de}(n)       & = & \mathsf{ue}(n) \cap \mathsf{pass}(n)\\
\mathsf{entrySeed}   & = & \mathsf{haltSeed}\ =\ \varnothing\\
\neg X               & = & \mathsf{allExprs} \setminus X\\
\textbf{Placement functions} & &\\[2pt]
\mathsf{availableOut}(\eta_a)(n)  & = & \mathsf{de}(n) \cup (\eta_a(n) \cap \mathsf{pass}(n))\\
\mathsf{earliest}(\pi_a,\eta_a)(n,n')   & = & \pi_a(n') \cap \neg\,\mathsf{availableOut}(\eta_a)(n) \cap (\,n{=}\mathsf{entry}\;?\;\mathsf{allExprs}\;:\;\neg\,\pi_a(n)\,)\\
\mathsf{latestNode}(\eta_p,\tau_p)(n)        & = & \eta_p(n) \cap (\mathsf{ue}(n) \cup \neg\,\tau_p(n))\\
\mathsf{latestEdge}(\pi_a,\eta_a,\eta_p)(n,n') & = & \big(\mathsf{earliest}(\pi_a,\eta_a)(n,n') \cup (\eta_p(n) \setminus \mathsf{ue}(n))\big) \setminus \eta_p(n')
\end{array}$
\caption{LCM node local sets, placement functions, and constants. The placement functions $\mathsf{earliest}$,
$\mathsf{latestNode}$, and $\mathsf{latestEdge}$ are parameterized by the foreign ghost
variables they read.}
\label{fig:lcm-locals}
\end{figure*}

% GENERATED by script/gsl2tex.py — do not edit by hand; regenerate via script/gen-gsl-figures.sh
% Requires \usepackage{listings}; reads the verbatim block from figures/lcm-spec.gsltex.
\ifdefined\gslListingsReady\else
  \lstdefinelanguage{GSL}{morekeywords={analysis,prophecy,history,predict,check,update,seed,within,when,meets},morecomment=[l]{--},sensitive=true}
  \lstdefinestyle{gslstyle}{language=GSL,basicstyle=\sffamily\small,keywordstyle=\bfseries,
    columns=fullflexible,keepspaces=true,extendedchars=true,showstringspaces=false,
    breaklines=true,breakatwhitespace=true,breakindent=6em,
    literate=
    {π}{{$\pi$}}1
    {η}{{$\eta$}}1
    {τ}{{$\tau$}}1
    {ₐ}{{${}_{a}$}}1
    {ₚ}{{${}_{p}$}}1
    {ᵤ}{{${}_{u}$}}1
    {∀}{{$\forall$}}1
    {→}{{$\to$}}1
    {∅}{{$\emptyset$}}1
    {∖}{{$\setminus$}}1
    {∩}{{$\cap$}}1
    {∪}{{$\cup$}}1
    {⊆}{{$\subseteq$}}1
    {∈}{{$\in$}}1
    {·}{{$\cdot$}}1
    {⇒}{{$\Rightarrow$}}1}
  \let\gslListingsReady\relax
\fi
\begin{figure*}[thbp]
\begin{lstlisting}[style=gslstyle]
analysis LCM {
  prophecy Anticipated πₐ : Assignments[Expr] {
    predict : ∀ n→n'. πₐ(n) ∖ ue(n) ⊆ πₐ(n')
    check   : ∀ n. πₐ(n) ⊆ ue(n) ∪ pass(n)
    seed    : πₐ(final) ⊆ haltSeed
    within  : ∀ n. πₐ(n) ⊆ allExprs
  }
  history Available ηₐ : Assignments[Expr] {
    update : ∀ n→n'. ηₐ(n') ⊆ de(n) ∪ (ηₐ(n) ∩ pass(n))
    seed   : ηₐ(entry) ⊆ entrySeed
    within : ∀ n. ηₐ(n) ⊆ allExprs
  }
  history Postponable ηₚ : Assignments[Expr] {
    update : ∀ n→n'. ηₚ(n') ⊆ earliest(πₐ,ηₐ)(n,n') ∪ (ηₚ(n) ∖ ue(n))
    seed   : ηₚ(entry) ⊆ entrySeed
    within : ∀ n. ηₚ(n) ⊆ allExprs
  }
  prophecy TauP τₚ : Assignments[Expr] {
    predict : ∀ n→n'. τₚ(n) ⊆ ηₚ(n')
    within  : ∀ n. τₚ(n) ⊆ allExprs
  }
  prophecy Used πᵤ : Assignments[Expr] {
    predict : ∀ n→n'. πᵤ(n') ⊆ πᵤ(n) ∪ ( latestNode(ηₚ,τₚ)(n') ∪ latestEdge(πₐ,ηₐ,ηₚ)(n,n') )
    check   : ∀ n. ue(n) ∖ latestNode(ηₚ,τₚ)(n) ⊆ πᵤ(n)
    within  : ∀ n. πᵤ(n) ⊆ allExprs
  }
  prophecy UsedOut τᵤ : Assignments[Expr] {
    predict : ∀ n→n'. πᵤ(n') ⊆ τᵤ(n)
    within  : ∀ n. τᵤ(n) ⊆ allExprs
  }
}
\end{lstlisting}
\caption{LCM analysis specification.}
\label{fig:lcm-spec}
\end{figure*}

\section{Lazy Code Motion (LCM)}
\label{sec:lcm}

Lazy Code Motion (LCM) moves expression evaluations to eliminate redundant and partially 
redundant expression evaluations. 
An expression placement is {\em safe} if it is 1) down safe --- each moved expression is inserted 
only at points where it is anticipated in the original program (evaluated in all execution traces
from the insertion point before its operands are redefined) --- and 2) use covering --- every use (computation) of the 
expression is preceded, on every path reaching it, by an insertion of the expression 
(so a computed value is always available when needed). LCM is also {\em computationally optimal} --- 
on every halting path a moved expression 
is evaluated at most as many times as for any other safe placement. Within computationally
optimal placements, LCM delivers a placement that minimizes the lifetimes of moved 
expressions. 

\subsection{LCM Transform}

The LCM transform inserts expression evaluations before nodes and on 
control flow edges in the original program, storing the value of the expression into a temporary 
variable, then replacing downstream computations of the expression with 
reads of the temporary variable.  Figure~\ref{fig:lcm-insert} presents the 
rules that specify where the insertions happen. $\mathsf{entry}(n)$ is 
the set of expression evaluations $t_e := e$ inserted before $n$ 
(here $t_e$ is the temporary that holds the value of $e$ for subsequent
downstream use).  $\mathsf{edge}(n,m)$ is the set of expression evaluations 
$t_e := e$ inserted on the edge between $n$ and $m$. The transform performs the
inserts and rewrites the original expression evaluations to either read a temporary
variable carrying a computed value of the expression (rule $\textsc{Replace}$) or keep
the original assignment (rule $\textsc{Keep}$).

The LCM analysis (Section~\ref{sec:lcm-analysis}) delivers the $\mathsf{insertBefore}(n)$, $\mathsf{insertEdge}(n,m)$,
and $\mathsf{recoverable}(n)$ sets that drive the transform. 
The insertion is determined by ghost variables $\pi_u(n)$ and $\tau_u(n)$ 
and sets $\mathsf{latestEdge}(\pi_a,\eta_a,\eta_p)(n,m)$ (the set of
expressions whose latest edge insertion is on the control flow edge from $n$ to $m$)
and $\mathsf{latestNode}(\eta_p,\tau_p)(n)$ (the set of expressions whose latest node insertion is
before $n$) (Figure~\ref{fig:lcm-locals}). The transform inserts an expression before
a latest node $n$ only if it cannot be inserted on a latest edge out of $n$.
The ghost variable $\tau_u(n)$ determines whether an inserted temp $t_e := e$ will 
be used downstream, preventing the insert if not. $\pi_u$ determines whether 
to replace an assignment $x:=e$ with an assignment $x:=t_e$ that reads a previously inserted
temp $t_e := e$ or leave the assignment $x:=e$ in place. 
$\mathsf{numbered}(e)$ identifies expressions $e$ that can be moved --- 
in our implementation all expressions $\odot\,a$ or $a \oplus b$ can be moved. 

We note that standard presentations of lazy code motion insert expression evaluations only before
and after nodes~\cite{DBLP:conf/pldi/KnoopRS92,DBLP:journals/toplas/KnoopRS94}. This approach works only if there are no {\em critical} edges
in the control flow graph --- i.e., if there are no edges from a node with multiple successors
to a node with multiple predecessors. Standard presentations accomplish this constraint by 
splitting critical edges before analyzing the program and applying the transformation, 
increasing the size of the program representation. Our version works for programs with
critical edges, does not split critical edges, and avoids this size increase. To the
best of our knowledge this paper presents the first correctness proof of this 
variant (machine checked or otherwise). Our version does require no self reads
(no assignment $x:=e$ with $x \in \mathsf{uses}(e)$) and implements a normalization
pass that uses temporary variables to rewrite and eliminate self read assignments. 

\subsection{LCM Analysis} 
\label{sec:lcm-analysis}

The analysis itself works with six interdependent ghost variables ---
four prophecy variables ($\pi_a$, $\pi_u$, $\tau_p$, and $\tau_u$) and two history variables
($\eta_a$ and $\eta_p$) (Figures~\ref{fig:lcm-spec} and \ref{fig:lcm-locals}).
We note that $\pi_u$ and $\tau_u$ exist only to prevent the insertion of 
{\em isolated} computations $t_e := e; x:=t_e$, where $x:=t_e$ is the only use of $t_e$.
If isolated computations are not a concern, a significant simplification is possible ---
drop $\pi_u$ and $\tau_u$, drop $\mathsf{recoverable}(n)$ from the preconditions 
of the $\textsc{Replace}$ and $\textsc{Keep}$ rules, and let 
$\mathsf{insertBefore}(n) = \mathsf{latestNode}(\eta_p,\tau_p)(n) \setminus \textstyle\bigcup_{m \in \mathsf{succ}(n)} \mathsf{latestEdge}(\pi_a,\eta_a,\eta_p)(n,m)$
and $\mathsf{insertEdge}(n,m) = \mathsf{latestEdge}(\pi_a,\eta_a,\eta_p)(n,m)$
in Figure~\ref{fig:lcm-insert} (see Figure~\ref{fig:lcm-insert-basic} in Section~\ref{sec:lcm-insert-basic}).

We call the two prophecy variables $\tau_p$ and $\tau_u$ {\em transfer variables}.
Transfer variables exist because the values of prophecy and history variables are defined only at 
node entry points. But LCM needs analysis results at node exit points. Transfer variables 
materialize exit point analysis results and make them available to the 
analysis and transform. The ghost variables work as follows:
\begin{itemize}
\item $\pi_a(n)$: A prophecy variable that predicts the set of expressions $e$
{\em anticipated} at $n$, i.e., the set of expressions $e$ that
are evaluated on all execution traces from $n$ before their operands
are redefined. The $\mathsf{predict}$ clause requires anticipated 
expressions $\pi_a(n')$ after $n$ to contain all anticipated expressions
$\pi_a(n)$ before $n$, except possibly expressions that
$n$ computes (the upwardly exposed expressions $\mathsf{ue}(n)$) --- these
upwardly exposed expressions may be the last occurrence in the trace before
the operands are redefined. The prediction may also include other expressions
that become available at $n'$. 
\[
\mathsf{predict}\; : \; \langle n,\_\rangle \to \langle n',\_\rangle\ \Rightarrow\ \pi_a(n) \setminus \mathsf{ue}(n) \subseteq \pi_a(n')
\]
The check clause requires $\pi_a(n)$ to include at most the upwardly exposed
expressions $\mathsf{ue}(n)$ and expressions $\mathsf{pass}(n)$ that
pass through $n$ (i.e., $n$ does not redefine their operands) ---
an expression can't be anticipated unless it is computed by $n$ or can be anticipated beyond $n$. 
\[
  \mathsf{check}\; : \;   \pi_a(n) \subseteq \mathsf{ue}(n) \cup \mathsf{pass}(n)
\]

\item $\eta_a(n)$: A history variable that records the {\em available} expressions. The $\mathsf{update}$
clause records that the expressions that $n$ evaluates (the downwardly exposed expressions $\mathsf{de}(n)$)
along with all expressions that are available at $n$ and pass through $n$ can be available at $n'$.

\[
\mathsf{update} \; : \; \langle n,\_\rangle \to \langle n',\_\rangle\ \Rightarrow\ \eta_a(n') \subseteq \mathsf{de}(n) \cup (\eta_a(n) \cap \mathsf{pass}(n))
\]

\item $\eta_p(n)$: A history variable that records the {\em postponable} expressions --- expressions
that can be evaluated sometime before $n$ but can be postponed until at least $n$. One of the goals of LCM is
to insert expression evaluations as late as possible to minimize live ranges. $\eta_p(n)$ identifies how
far expression evaluations can be postponed in pursuit of this goal. An expression $e$ can be postponed until at least $n'$
if either 1) the edge from $n$ to $n'$ is its earliest insertion point or 2) it can be postponed until $n$
and $n$ does not evaluate the expression (if $n$ does evaluate $e$ it must be inserted before $n$ and not
postponed any further). 

\[
  \mathsf{update} \; : \; \langle n,\_\rangle \to \langle n',\_\rangle\ \Rightarrow\ \eta_p(n') \subseteq \mathsf{earliest}(\pi_a,\eta_a)(n,n') \cup (\eta_p(n) \setminus \mathsf{ue}(n))
\]

Conceptually, $\mathsf{earliest}(\pi_a,\eta_a)(n,n')$ contains expressions whose earliest possible edge insertion
is on the edge from $n$ to $n'$ (Figure~\ref{fig:lcm-locals}). It imposes three constraints on expressions
$e \in \mathsf{earliest}(\pi_a,\eta_a)(n,n')$: 1) $e$ must be anticipated at $n'$ ($e \in \pi_a(n')$), 2) $e$ is not already available
out of $n$ (it would be available out of $n$ if $e \in \mathsf{de}(n)$ or $e \in \eta_a(n) \cap \mathsf{pass}(n)$), and
3) $e$ is not anticipated at $n$ ($e \notin \pi_a(n)$) --- if $e$ were anticipated at the entry of $n$ it could be
inserted on an edge into $n$, so its earliest insertion would not be on an edge out of $n$. Note that $e$ can be
not anticipated at $n$ but inserted on an edge out of $n$ when $n$ has multiple successors and $e$ is anticipated
on only some of them: $\pi_a(n)$ requires that $e$ be anticipated at \emph{all} successors of $n$ (or computed by $n$).

\item $\tau_p(n)$: A transfer variable that identifies the expressions postponable until the
exit of $n$ --- conceptually the history variable $\eta_p$ at the exit of $n$ instead of the entry of $n$. 
The $\mathsf{predict}$ clause requires $\tau_p(n)$ to be contained in the postponable set $\eta_p(n')$:
\[
\mathsf{predict}\ \langle n,\_\rangle \to \langle n',\_\rangle\ \Rightarrow\ \tau_p(n) \subseteq \eta_p(n')
\]
The $\mathsf{latestNode}(\eta_p,\tau_p)(n)$ placement function (Figure~\ref{fig:lcm-locals}) reads $\tau_p$ to force an
expression to be evaluated at $n$ rather than postponed further whenever it is either used at $n$ or not
postponable at the exit of $n$ ($\neg\,\tau_p(n)$). An expression $e$ that is not used at $n$ 
can be postponable at $n$ but not at the exit of $n$ when it is not used
along one of the paths out of $n$ and is therefore not postponable at one of the successors of $n$. 
\[
\mathsf{latestNode}(\eta_p,\tau_p)(n) = \eta_p(n) \cap (\mathsf{ue}(n) \cup \neg\,\tau_p(n))
\]

\item $\pi_u(n)$: A prophecy variable that predicts the set of expressions $e$ whose temporary
$t_e$ is {\em recoverable} at $n$ --- i.e., $t_e := e$ is inserted at or before $n$ and $t_e$ is read at or beyond $n$ 
by a recomputation of $e$ distinct from where $t_e$ is inserted. A complexity is preventing 
{\em isolated} computations $t_e := e; x:=t_e$, where $x:=t_e$ is the only use of $t_e$ --- in this case
the transform leaves the original $x := e$ in place. The $\mathsf{check}$ clause enables the elimination
of isolated computations by discarding expressions in $\mathsf{latestNode}(\eta_p,\tau_p)(n)$ from the set of expressions $\mathsf{ue}(n)$ that $n$ evaluates:
\[
\mathsf{check}\; : \; \mathsf{ue}(n) \setminus \mathsf{latestNode}(\eta_p,\tau_p)(n) \subseteq \pi_u(n)
\]
The $\mathsf{predict}$ clause predicts recoverable temps at $n'$ --- $t_e$ is
recoverable at $n'$ only if it was 1) recoverable at $n$ ($\pi_u(n)$), 2) inserted before 
$n'$ ($\mathsf{latestNode}(\eta_p,\tau_p)(n')$), or 3) inserted
on the edge into $n'$ ($\mathsf{latestEdge}(\pi_a,\eta_a,\eta_p)(n,n')$):
\[
\begin{aligned}
\mathsf{predict} \; : \; \langle n,\_\rangle \to \langle n',\_\rangle\ \Rightarrow\ \pi_u(n') \subseteq{}
  & \pi_u(n) \cup \mathsf{latestNode}(\eta_p,\tau_p)(n') \\
  & {}\cup \mathsf{latestEdge}(\pi_a,\eta_a,\eta_p)(n,n')
\end{aligned}
\]
Note that $\mathsf{latestNode}(\eta_p,\tau_p)(n')$ is included in the $\mathsf{predict}$ clause but excluded from the $\mathsf{check}$ clause ---
the $\mathsf{predict}$ clause establishes an upper bound on $\pi_u(n')$ while the $\mathsf{check}$ clause imposes a lower
bound. Together, these clauses enable these expressions 
to correctly appear in $\pi_u(n')$ if they are used downstream of $n'$ while not requiring them to appear in 
$\pi_u(n')$ if they are not. 

\item $\tau_u(n)$: A transfer variable that identifies the expressions recoverable at some
successor of $n$ --- conceptually the prophecy variable $\pi_u(n)$ at the exit of $n$
instead of the entry of $n$ (prophecy and history variables are defined at node entry points). 
Its $\mathsf{predict}$ clause requires $\tau_u(n)$ to contain every expression in the recoverable
set $\pi_u(n')$ of all successors $n'$ of $n$:
\[
\mathsf{predict}\ \langle n,\_\rangle \to \langle n',\_\rangle\ \Rightarrow\ \pi_u(n') \subseteq \tau_u(n)
\]
The transform (Figure~\ref{fig:lcm-insert}) reads $\tau_u$ to ensure that it inserts assignments $t_e := e$ 
only where the temp is recoverable downstream so an isolated computation is never inserted:
\[
\begin{aligned}
\mathsf{insertBefore}(n) &= \bigl(\mathsf{latestNode}(\eta_p,\tau_p)(n) \cap \tau_u(n)\bigr) \\
  &\quad {}\setminus \textstyle\bigcup_{m \in \mathsf{succ}(n)} \mathsf{latestEdge}(\pi_a,\eta_a,\eta_p)(n,m) \\
\mathsf{insertEdge}(n,m) &= \mathsf{latestEdge}(\pi_a,\eta_a,\eta_p)(n,m) \cap \tau_u(n)
\end{aligned}
\]

\end{itemize}

The LCM analysis highlights how prophecy and history variables, along with transfer variables, can support
the specification and verified use of even quite complex analyses with multiple interdependent ghost variables. 

\subsection{LCM Correctness} 

The LCM correctness theorem states that if the original program halts, then
the transformed program also halts and the values of the observable variables
are the same in both executions. LCM also preserves faults --- if the
original program faults (divide or mod by zero), then so does the transformed
program. The transform does not, however, preserve divergence --- the transform
may insert a faulting command in front of an infinite loop, turning a divergent 
execution into a fault.

% transform_preserves_halt
% transform_preserves_faulting

The LCM correctness proof is a forward simulation over the original program $P$
and transformed program $T$. As in the PDCE proof, it matches each step
$\langle n_o,\sigma_o\rangle \to \langle n_o',\sigma_o'\rangle$ from the original
program against potentially multiple steps
$\langle n_t,\sigma_t\rangle \to \cdots \to \langle n_t',\sigma_t'\rangle$
from the transformed program. The transform replaces each original node assignment 
or $\mathsf{noop}$ node $n$ by a block 
consisting of a (potentially empty) {\em entry chain} of 
assignments $t_e := e$ inserted before $n$, the (possibly rewritten) original command,
and a (potentially empty) {\em exit chain} of assignments $t_e := e$ inserted on $n$'s
outgoing edge. 
An $\mathsf{ifz}\ x$ node $n$ has multiple successors and therefore no single outgoing
edge. Its block therefore consists of the entry chain, the branch itself, followed by 
{\em edge chains} --- one per outgoing successor edge --- each containing
the assignments $t_e := e$ inserted on that outgoing edge. 

The proof maintains an invariant relating a set $M$
of expressions $e$ currently \emph{in flight} --- the assignment $t_e := e$
is already inserted but not yet read by all downstream uses --- to values in 
the two stores $\sigma_o$ and $\sigma_t$. $M$ is not a direct function of
the ghost variables but is instead a history set threaded along the trace. 
At corresponding configurations the invariant
$\mathsf{Match}\,\langle n_o,\sigma_o\rangle\,\langle n_t,\sigma_t\rangle\,M$ holds:
\[
\begin{array}{l@{\qquad}l}
    n_t = \mathsf{block}(n_o), & \\[4pt]
    \mathsf{eval}\,\sigma_t\,t_e \;=\; \mathsf{eval}\,\sigma_o\,e
      & \text{for every } e \in M,\\[4pt]
    \sigma_t\,x \;=\; \sigma_o\,x
      & \text{for every \emph{non-fresh} } x\ \ (\forall e.\ t_e \neq x),\\[4pt]
    M \subseteq \mathsf{allExprs}. &
\end{array}
\]
i.e., the two stores agree on every variable except the fresh temporaries $t_e$, and for each
expression $e$ in flight its temporary $t_e$ evaluates in $\sigma_t$ to the value $e$ evaluates to in
$\sigma_o$. A variable is \emph{non-fresh} when it is none of the introduced temporaries --- this covers the
original source variables \emph{and} any temporaries from earlier passes.

The invariant is accompanied by a \emph{coverage} side condition over the ghost variables,
\[
  \mathsf{Cov}(n_o, M) \;\triangleq\;
  \bigl(\mathsf{used}(n_o) \setminus \mathsf{entry}(n_o)\bigr) \setminus \mathsf{out}(n_o)
  \;\subseteq\; M,
\]
i.e., every expression the analysis marks as \emph{used} at $n_o$ whose temp $t_e := e$ is not inserted at
$n_o$'s entry ($\mathsf{entry}(n_o)$) or on an outgoing edge ($\mathsf{out}(n_o)$) is already in flight ---
which is what lets the transform replace an original computation $x := e$ by the copy $x := t_e$.

The core simulation lemma preserves both the mapping and coverage across one step of the original
program $P$, with the in-flight set updated according to the expressions inserted at the entry node
and edge of the step:
\[
  \mathsf{step}(n_o, n_o', M) \;\triangleq\;
  \bigl((M \cup \mathsf{entry}(n_o)) \cap \mathsf{pass}(n_o)\bigr)
  \,\cup\, \mathsf{exit}(n_o) \,\cup\, \mathsf{edge}(n_o, n_o'),
\]

%\[
%\begin{array}{l}
%  \mathsf{Match}\,\langle n_o,\sigma_o\rangle\,\langle n_t,\sigma_t\rangle\,M
%  \;\wedge\; \mathsf{Cov}(n_o,M)
%  \;\wedge\; \mathsf{nofault}(n_o,\sigma_o)
%  \;\wedge\; \langle n_o,\sigma_o\rangle \to \langle n_o',\sigma_o'\rangle \\[4pt]
%  \quad\Longrightarrow\;
%  \exists\,\sigma_t'.\;
%  \langle n_t,\sigma_t\rangle \to^{*} \langle \mathsf{block}(n_o'),\sigma_t'\rangle
%  \;\wedge\;
%  \mathsf{Match}\,\langle n_o',\sigma_o'\rangle\,\langle \mathsf{block}(n_o'),\sigma_t'\rangle\,\mathsf{step}(n_o,n_o',M),
%\end{array}
%\]
% together with coverage maintenance $\mathsf{Cov}(n_o,M) \Rightarrow \mathsf{Cov}(n_o',\mathsf{step}(n_o,n_o',M))$.

A case analysis on the command at $n_o$ (noop, conditional, or assignment) discharges the lemma, using the
ghost conditions to align the step with the corresponding sequence in the transformed program.
Here $\mathsf{nofault}(n_o,\sigma_o)$ requires each inserted expression to evaluate without faulting
($\mathsf{eval}\,\sigma_o\,e \neq \mathsf{none}$ for $e$ in $n_o$'s entry/exit/edge chains); on a halting run
this follows from down safety, which is why --- in contrast to PDCE --- a \emph{valid} analysis result $G$
is not sufficient for LCM correctness: the proof additionally requires that $G$ be \emph{extremal}
($\mathsf{Extremal}(G)$). The pair $(\mathsf{Match},\mathsf{Cov})$ is therefore the invariant --- established
at the entry ($M = \varnothing$, equal stores) and proved across corresponding steps in the original
and transformed statements, with the ghost variables enabling the proof of the key step lemma. 

\subsection{LCM Optimality} 

The LCM optimality theorems characterize properties of the transformed 
program $T = \mathsf{transform}\; P\; G$ under the extremal analysis result $G$. 
We consider two kinds of optimality --- {\em computational optimality} (how many times the program evaluates an expression)
and {\em lifetime optimality} (live ranges of temporaries $t_e$). We prove computational optimality against the
space of safe placements of expression evaluations $t_e := e$ in the original program $P$ --- placements that compute each expression only
where it is down safe and use covering, with down safety characterized by 
the extremal value of the prophecy variable $\pi_a$ (anticipated) and use covering characterized by
the extremal value of the history variable $\eta_a$ (available). We prove lifetime optimality against the space of all computationally
optimal placements of expression evaluations $t_e := e$ in the original program $P$. 
\begin{itemize}
\item {\bf Computational Optimality:} On every halting trace of $P$, 
$T$ evaluates each expression $e$ no more often than any safe placement does. 
Each safe placement recomputes $t_e := e$ at least once per point where $e$ becomes newly anticipated and 
unavailable (an \emph{earliest} insertion point), while $T$ evaluates $e$ at most once per such insertion point. Therefore
$T$ is computationally optimal. 

% transform_evalCount_le_safe

\item {\bf Lifetime Optimality:} Computationally optimal placements are all driven by analysis results $G'$ with
extremal $\eta_a$ and $\pi_a$
but may have only valid $\eta_p$ (postponable) (and therefore only valid $\tau_p$ as well).
The other ghost variables ($\pi_u$ and $\tau_u$) are used only to prevent isolated expressions in the 
transformed program. We therefore compare against placements driven by extremal $\eta_a$ and $\pi_a$ 
and valid $\eta_p$. Earliest expression insertion points are driven by $\eta_a$ and $\pi_a$ (and are therefore
the same for all computationally optimal placements). Latest expression insertion points are driven by
$\eta_p$ --- because any valid $\eta_p$ is a subset of the extremal $\eta_p$, extremal $\eta_p$ 
placements postpone expression evaluation at least as much as valid $\eta_p$ and therefore minimize live ranges. 

% pathLiveLen_le (built on liveRegion_minimal)

\end{itemize}

\section{Nexis Analysis Specification Language and Implementation} 
\label{sec:nexis}

Nexis analysis specifications are written in a domain specific ghost specification 
language (Figures~\ref{fig:pdce-spec} and~\ref{fig:lcm-spec}). These specifications
use node local functions, placement functions, and constants written in Lean 4
(Figures~\ref{fig:pdce-locals} and~\ref{fig:lcm-locals}). The base
language implementation (Figure~\ref{fig:ir-syntax}) provides a range of node local functions
that ghost specifications can build on. Developers can use these node local functions,
define their own, or use some combination of the two.

\subsection{Node Local Sets, Placement Functions, and Constants}

Node local sets, placement functions, and constants are defined 
in a plain Lean file written against an API over base language
elements --- configurations, programs, commands, variables, and expressions. Each Lean def
denotes a node local set (e.g.\ $\mathsf{born}(n)$, $\mathsf{pass}(n)$, $\mathsf{kills}(n)$ 
in Figure~\ref{fig:pdce-locals}), a boolean helper, a placement function, or a
whole program constant, built from the domain operations $\cup$, $\cap$, and
$\setminus$ together with a union
$\bigcup_n$ over nodes $n$ and grounded out on the base language API and 
Lean constructs. Placement functions may additionally reference the prophecy
and history variables (see $\mathsf{latestNode}(\eta_p, \tau_p)(n)$ and 
$\mathsf{latestEdge}(\pi_a, \eta_a, \eta_p)(n,n')$ in Figure~\ref{fig:lcm-spec}).
Placement functions can be either {\em node} functions that are a function of a 
single node $n$ or {\em edge} functions that are a function of two nodes $n$ and $n'$.

\subsection{Ghost Specification Language}

The ghost specification language defines each analysis as a sequence of ghost variables. 
Each ghost variable is either a $\mathsf{prophecy}$ or $\mathsf{history}$ variable. 
A ghost variable is declared as follows (Figure~\ref{fig:pdce-spec}):
\[
  \mathbf{prophecy}\ \ \mathsf{Live}\ \ \pi \,:\, \mathsf{Variables}[\mathsf{Var}]
\]
The declaration specifies whether the variable is a prophecy or history variable, 
the name of the ghost variable ($\mathsf{Live}$), the ghost
variable itself ($\pi$), the name of the domain of the ghost variable ($\mathsf{Variables}$),
and the elements of the sets of the domain ($\mathsf{Var}$) from the base language. So
in the declaration above, $\mathsf{Live}$ prophecy variables are sets of program variables. 

\subsubsection{Generated Predicates}

The implementation generates a predicate that is true for ghost variable values that 
satisfy the ghost variable set inclusion constraints as specified by the clauses in the 
ghost variable specification ($\mathsf{Live}\; P \; \pi$). The predicate is a Lean structure 
whose fields restate the clauses of the ghost variable declaration. The name of 
this structure is the ghost variable name. 

In general analyses may include multiple ghost variables. The implementation generates
a {\em valid} predicate that is the conjunction of the ghost variable predicates. It also
generates an {\em extremal} predicate that is true when all of the ghost variable values are
the least or greatest (under subset inclusion, with the direction of the inclusion 
inferred by the implementation) among all valid ghost variable values. 

Transforms and their proofs of correctness and optimality are universally quantified 
over valid (and potentially also extremal) analyses as specified by these predicates. 
The current Nexis implementation generates code that invokes a worklist based
dataflow solver to obtain a valid and extremal value for each ghost variable. 
The generated predicates therefore serve as an 
abstraction boundary between the transform and the dataflow algorithm. 

\subsubsection{Generated Files and Theorems}
In the current implementation each analysis is specified in a single file that is parsed
and processed into three generated Lean files. The first file contains the connection to the dataflow solvers,
including the extracted transfer functions, solver invocations, and relevant theorems (for example,
the validity and extremality of the dataflow result with respect to the transfer functions). 
The second contains the restatements of the ghost variable clauses, theorems that prove 
that the dataflow analysis results are valid and extremal solutions to the inclusion
constraints as specified by the ghost variable clauses,
and the analysis result bundle of ghost variables that the transforms consume. The third contains
the progress and preservation theorems.

The translation from the file containing the ghost specifications into the three Lean files
is currently unverified. The restated ghost variable clauses in the second generated Lean file are, however, 
legible, inspectable, and fully verified.
If a fully verified system is a goal, it is straightforward to lift this restatement to 
be the definitive ghost variable specification. 

\subsubsection{$\mathsf{predict}$ Clauses}

A predict clause identifies an inclusion relation that (typically nondeterministically)
predicts the value of a prophecy variable at a node $n$ or $n'$ given the value at an
adjacent node. The natural formulation is forward --- it predicts the value $\pi(n')$ 
at a next node $n'$ given the value $\pi(n)$ at a previous node --- but the language 
also supports specifying $\pi(n)$ given the value at $\pi(n')$. 
The prophecy variable appears isolated on one side of the inclusion 
with the other side a set expression in the following grammar:
\[
  \mathit{Predict} \;\Coloneqq\;
  \mathbf{predict}\;\mathord{:}\;\; \forall\, n\!\to\!n'.\;\;
  \pi(\nu) \subseteq E \;\;\bigm|\;\; E \subseteq \pi(\nu),
  \qquad \nu \in \{\,n,\;n'\,\}.
\]
\[
  E \;\Coloneqq\; r \;\bigm|\; \emptyset \;\bigm|\;
  E \cup E \;\bigm|\; E \cap E \;\bigm|\; E \setminus E \;\bigm|\; (\,E\,),
  \qquad
  r \;\Coloneqq\; f \, [\,(\,\overline{\gamma}\,)\,] \, [\,(\,\overline{\nu}\,)\,],
\]
Here $\cap$ and $\setminus$ bind more tightly than $\cup$ and all operators are
left-associative. A leaf reference $r = f(\overline{\gamma})(\overline{\nu})$ names a
node local function, a foreign ghost (a previously defined ghost variable, 
see Figure~\ref{fig:lcm-spec}), or a placement function: $f$ is the name, the optional
$\overline{\gamma}$ are the foreign ghosts a placement reads, and the node arguments
$\overline{\nu}$ are drawn from $\{n,n'\}$ (a single node $n$ or an edge pair $(n,n')$). The
prophecy variable itself may appear among these leaves as $\pi(n)$ or $\pi(n')$. 

The clause must satisfy several conditions.  First, the prophecy variable 
must occur \emph{monotonically}: it may appear as an operand of $\cup$ or $\cap$, or as the
left hand side of a difference ($\pi(\nu)\setminus X$), but never on the right hand side
of the difference or otherwise negated. This condition ensures that the extracted transfer function 
is monotone and its extremal fixpoint is well defined. 

When the nonisolated side is a \emph{single bare foreign ghost} with no occurrence of the ghost 
variable itself --- for example, $\tau(n) \subseteq \eta(n')$ --- the 
clause has no recursive self dependence and defines a transfer variable that makes 
the analysis value at node exit points nameable --- for example, 
$\forall n \to n' . \tau(n) \subseteq \eta(n')$ makes $\eta$ at the exit point of $n$ nameable 
as $\tau(n)$.

\subsubsection{$\mathsf{check}$ Clauses} 

% Requires: amsmath, mathtools (\Coloneqq, \bigm|). Metavariables match the predict-clause text:
% π the prophecy variable, n/n' the endpoints of an edge, e the set expression, z the contributed
% element, y an existentially bound element, f a node local relation on elements.

A check clause specifies a condition that a prophecy variable must satisfy at a node or 
step. A \emph{clamp} clause specifies either a lower bound $\forall n . E \subseteq \pi(n)$ 
or an upper bound $\forall n. \pi(n) \subseteq E$, where $E$ is a set expression as above. 
A \emph{guard} clause holds at a node only when a guard on the value at the
adjacent node holds --- the guard tests the value $\pi(n')$ at the adjacent node, and when it
is satisfied the check requires the value $\pi(n)$ to contain a result. As with a $\mathsf{predict}$ clause the
prophecy variable appears isolated on one side of the inclusion. Clamp clauses quantify over
single nodes. Guard clauses quantify over edges:
\[
  \begin{array}{r@{\;}c@{\;}l}
    \mathit{Check}
      &\Coloneqq& \mathbf{check}\;\mathord{:}\;\; \forall\, n.\;\;
                  \bigl( E \subseteq \pi(n) \bigm| \pi(n) \subseteq E \bigr) \\[3pt]
      &\big|&     \mathbf{check}\;\mathord{:}\;\; \forall\, n\!\to\!n'.\;\;
                  \bigl( E \subseteq \pi(n) \bigm| z \in \pi(n) \bigr) \;\mathbf{when}\; g
  \end{array}
\]
The guard $g$ is a monotone combination of atoms that test the incoming value:
\[
  \begin{array}{r@{\;}c@{\;}l}
    g &\Coloneqq& a \bigm| g \wedge g \bigm| g \vee g \bigm| (\,g\,) \\[3pt]
    a &\Coloneqq& E \;\mathbf{meets}\; \pi(n') \bigm| E \subseteq \pi(n') \\[3pt]
      &\big|&     \exists\, y \in \pi(n').\; f(y,z) \\[3pt]
      &\big|&     z \in \pi(n') \bigm| z \in E
  \end{array}
\]
where $\wedge$ is conjunction (and), $\vee$ is disjunction (or), 
$\wedge$ binds more tightly than $\vee$, $E$ is a set expression as above, $z$
is the element the guarded check contributes and $y$ an existentially bound element, and $f$
is a node local relation on elements (optionally node-indexed, $f(\overline{\nu})(y,z)$). 
$\pi(n')$ is the value of the prophecy variable at the next node. The final atom
$z \in E$ is the one form that does not read $\pi(n')$. 

Check clause guards must be \emph{monotone} in $\pi(n')$: $\pi(n')$ may appear only on the
right hand side of $\mathbf{meets}$, $\subseteq$, $\in$, or $\exists\,\cdot\in$, never negated or on the right
hand side of a difference. This condition keeps the clause monotone so that it can be combined
with the $\mathsf{predict}$ clause to deliver an extremal solution to the prophecy variable
constraints. The $z \in \pi(n') \;\mathbf{when}\; g$ form can express any monotone condition
that a node must satisfy given the value at an adjacent node. 

\subsubsection{$\mathsf{update}$ Clauses}

An $\mathsf{update}$ clause updates the value of a history variable at a node given the value at an
adjacent node. As with a $\mathsf{predict}$ clause the history
variable appears isolated on one side of the inclusion, with the other side a set expression $E$:
\[
  \mathit{Update} \;\Coloneqq\;
  \mathbf{update}\;\mathord{:}\;\; \forall\, n\!\to\!n'.\;\;
  \eta(\nu) \subseteq E \;\;\bigm|\;\; E \subseteq \eta(\nu),
  \qquad \nu \in \{\,n,\;n'\,\},
\]
Here $E$ and the monotonicity condition are the same as for predict clauses. The natural
formulation updates the value $\eta(n')$ at the next node $n'$ given the value $\eta(n)$ at the
previous node, but the language supports specifying $\eta(n)$ given $\eta(n')$. 
As with a predict clause, when the nonisolated side is a single bare foreign ghost, 
the clause defines a transfer variable that makes analysis results at node exit
points nameable. 

\subsubsection{$\mathsf{always}$ Clauses}
An $\mathsf{always}$ clause specifies a condition on a history variable $\eta$. As for 
$\mathsf{check}$ clauses, there are two forms of $\mathsf{always}$ clauses --- clamp 
clauses and guard clauses. Clamp clauses impose lower or upper bounds as for $\mathsf{check}$
clauses.  Guard clauses constrain $\eta(n')$:
\[
  \begin{array}{r@{\;}c@{\;}l}
    \mathit{Always}
      &\Coloneqq& \mathbf{always}\;\mathord{:}\;\; \forall\, n.\;\;
                  \bigl( E \subseteq \eta(n) \bigm| \eta(n) \subseteq E \bigr) \\[3pt]
      &\big|&     \mathbf{always}\;\mathord{:}\;\; \forall\, n\!\to\!n'.\;\;
                  \bigl( E \subseteq \eta(n') \bigm| z \in \eta(n') \bigr) \;\mathbf{when}\; g
  \end{array}
\]
The guard $g$, its atoms, and the monotonicity condition are exactly as for a 
$\mathsf{check}$ clause, with $\eta(n)$ in place of $\pi(n')$. 

\subsection{$\mathsf{seed}$ and $\mathsf{within}$ Clauses}

A $\mathsf{seed}$ clause specifies the values of ghost variables at program entry or halt points. 
A $\mathsf{within}$ clause specifies that the values of ghost variables stay within 
a finite universe of values such as all variables, assignments, or expressions in the program. 
The generator uses $\mathsf{within}$ clauses to fix the width of the bit vectors in the 
generated dataflow problem.

\subsection{Dataflow Backend}
\label{sec:dataflow}

The Nexis generator processes the analysis specification to emit, for each ghost variable, a dataflow
problem together with proofs that its solution is valid and extremal.

\subsubsection{Dataflow Universe}
In the current Nexis implementation every analysis domain is a finite set, implemented as a HashSet and encoded as
a fixed width bitvector. The scope is therefore ghost variables over finite sets of items, with the
constraints implemented as finite bitvector dataflow problems. For each ghost variable
the dataflow generator reads the dataflow universe $U$ off the $\mathsf{within}$ clause. It assigns each element
in the universe a position within a bitvector of length $|U|$ (example universes include all variables,
assignments, or expressions in the program) and runs the analysis over bitvectors of length $|U|$.
Set union, intersection, difference, and inclusion are implemented as bitwise operations. 
The exposed solution and every specification are stated over the
corresponding finite sets, so a transform and its correctness and optimality proofs never mention a
bitvector or any other property specific to the dataflow implementation. 

\subsubsection{Direction and Confluence Operator}

The current implementation generates four kinds of dataflow problems, one for each 
combination of direction (forward or backward) and confluence operator (union or intersection). 
Prophecy variables are implemented with backward analyses, history variables with forward
analyses. The confluence operator is determined by the {\em inclusion direction} of the ghost variable ---
prophecy variables of the form $E \subseteq \pi(n')$ (for example, 
the anticipated prophecy variable $\pi_a$ in Figure~\ref{fig:lcm-spec})
use intersection ($\cap$), prophecy variables of the form $\pi(n') \subseteq E$ 
(for example, the live prophecy variable $\pi$ in Figure~\ref{fig:pdce-spec}) use union ($\cup$),
history variables of the form $E \subseteq \eta(n')$ use union, and 
history variables of the form $\eta(n') \subseteq E$ use intersection. 

Transfer variables are implemented with a confluence operator applied to an
already computed ghost variable at adjacent nodes --- for example,
$\tau_p(n)$ in Figure~\ref{fig:lcm-spec} is computed as the
intersection ($\cap$) of $\eta_p(n')$ over all successors $n'$ of $n$.

\subsubsection{Dataflow Transfer Functions}

% Any monotone transfer function is expressible in the Nexis clause language. The Nexis 
% implementation processes the clauses to generate dataflow transfer functions

The natural form of a history variable $\mathsf{update}$ 
clause reads directly as a forward transfer --- 
it specifies the successor value $\eta(n')$ in terms of the current value $\eta(n)$.
The natural form of a prophecy variable $\mathsf{predict}$ clause also reads forward,
but is implemented with a backward analysis. For prophecy variables
the generator therefore inverts the inclusion to obtain a form that specifies the
current value $\pi(n)$ in terms of the successor value $\pi(n')$:
\[
  \begin{array}{r@{\;}c@{\;}l}
    \pi(n') \subseteq \pi(n) \cup \mathit{f}(n)
      &\Longrightarrow& \pi(n') \setminus \mathit{f}(n) \subseteq \pi(n) \\
    \pi(n) \setminus \mathit{f}(n) \subseteq \pi(n')
      &\Longrightarrow& \pi(n) \subseteq \pi(n') \cup \mathit{f}(n)
  \end{array}
\]
The resulting backward transfer reads directly off the inverted constraint. 
For prophecy variable analyses whose $\mathsf{predict}$ clauses do not
fit this form Nexis supports $\mathsf{predict}$ clauses 
that isolate $\pi(n)$ to express $\pi(n)$ in terms of the successor 
value $\pi(n')$. 

$\mathsf{check}$ and $\mathsf{always}$ clauses extend the analysis as follows. Clamp 
clauses bound the ghost variable at each node --- a lower bound requires the value 
of the ghost variable to include a given set, an upper bound requires a given set to 
include the value of the ghost variable. The generated transfer functions include the 
clamp clause constraints, which are applied as the last step of the computation of the
new prophecy variable value $\pi(n)$ every time the value is computed.

Guard clauses add elements conditioned on incoming dataflow values --- for example, 
the guard clause (Figure~\ref{fig:pdce-spec}):
\[
  \mathsf{check} : \langle n,\_\rangle \to \langle n',\_\rangle \;\Rightarrow\;
  \mathsf{rhsVars}(n) \subseteq \pi(n)\ \ \mathbf{when}\ \ \mathsf{defVars}(n) \cap \pi(n') \neq \varnothing
\]
adds $\mathsf{rhsVars}(n)$ to $\pi(n)$ when $\mathsf{defVars}(n) \cap \pi(n') \neq \varnothing$. The
computation proceeds as follows: 1) the transfer function from the $\mathsf{predict}$ clause is
applied to each incoming prophecy variable $\pi(n')$, 2) the guard clause is evaluated to obtain
a set of elements, which 3) are added to the set of elements produced by the transfer function to
obtain a result for each successor of the current node $n$. The confluence operator (union or
intersection) is 4) applied to the successor results and 5) the clamp clause constraints are
applied to derive the new prophecy variable value at the current node $n$. 

\subsubsection{Solvers and Proofs}
Nexis comes with a worklist dataflow solver and a reference fixpoint engine. The fixpoint engine
is an inefficient solver used in proofs involving the dataflow analysis results. A proof that
the worklist dataflow solver and the fixpoint engine compute the same result enables the 
implementation to soundly use the more efficient worklist solver. The worklist dataflow solver
is configured to provide four solution modalities, one for each 
combination of direction (forward or backward) and confluence operator (union or intersection).

For each ghost variable the generator emits a 
theorem that lifts the solver validity and extremality proofs --- stated over the
transfer calculus and bitvectors --- into validity and extremality proofs of the 
ghost-variable constraints stated over the operational semantics and finite sets. 
Nexis supports analyses (such as LCM) with multiple interdependent ghost variables. 
For these analyses Nexis generates a solver that computes the ghost variable
values in dependence order so that each ghost solve is built on the correct values of the 
foreign ghosts that it reads. This solver is proved to deliver 
ghost variable values that satisfy the required validity and extremality properties. 

The current system has no $\mathsf{sorry}$ed proofs, and every emitted theorem is checked 
to use only the standard axioms $\mathsf{propext}$, $\mathsf{Classical.choice}$, and
$\mathsf{Quot.sound}$.

\section{Related Work}
\label{sec:related}

In this paper we formalize and implement prophecy variables to enable forward reasoning, 
specifically forward simulation proofs, over properties that involve
the future execution of the program. Here we contrast with 
the recent use of prophecy variables for
program verification~\cite{JungLPRTDJ20,Vafeiadis08,ZhangFFSL12} as
well as the traditional use of prophecy variables for proving forward simulation
relations between state machines~\cite{AbadiL91,LynchV95,LynchDistributedAlgorithms}.
An (unpublished in any peer reviewed venue) manuscript introduces prophecy variables to
enable forward reasoning about program analysis properties that involve the future execution of
the program~\cite{anonymous2020}. The manuscript uses prophecy variables to specify
two program analysis problems, live variables and very busy expressions, that require information
about the future execution of the program and uses a backward dataflow analysis to obtain correct
prophecy variable predictions. There is no implementation and none of the proofs are machine
checked. Prophecy variables have recently been used to apply staged optimizations to programs written
in a domain specific language embedded in C++, with repeated program executions delivering
sound prophecy variable predictions~\cite{brahmakshatriya2026backwardsdataflowanalysisusing}. 

% the program~\cite{rinard2020dataflowanalysisprophecyhistory}. 

Cobalt enables compiler developers to specify a range of dataflow optimizations
(such as constant propagation and partial dead assignment elimination)~\cite{LernerMC03}.
Cobalt uses temporal logic predicates over the control flow graph (CFG), with 
each optimization specified by a transformation pattern whose guard specifies
a condition over sequences of actions in paths in the control flow graph that 
must hold for the transformation to be legal. Cobalt has separate constructs for specifying forward and backward
optimizations --- forward guards reason about forward properties, backward guards
reason about backward properties. Our research differs in that 1) it works with individual transitions
in the operational semantics, not global properties of static program representations and 
2) it uses prophecy variables (a forward concept) to enable forward simulation proofs, 
not backward guards that reason about backward properties. 

Rhodium was developed to eliminate the complexity of using temporal logic formulas and implicit dataflow facts
to reason about paths in control flow graphs~\cite{LernerMRC05}. Like
Cobalt, Rhodium operates over a static control flow graph representation. 
Rhodium uses de facto abstraction functions (expressed as
predicates over concrete program states) to establish the connection
between concrete program states and dataflow facts and state extensions
(a form of instrumented semantics) to support analyses that extract information
about the past execution of the program not present in standard concrete program
states. Like Cobalt, Rhodium has separate support for forward and backward analyses;
subsequent work on automatically inferring correct propagation rules supports only
forward rules~\cite{ScherpelzLC07}. Rhodium operates within the standard dataflow
analysis paradigm, including abstract interpretation and explicit control flow
merge functions. Our research differs in that 1) it works with individual transitions
in the operational semantics, not static program representations, 2) it 
uses prophecy variables (a forward concept) to enable forward simulation proofs, 
not backward analyses, 3) it discards the entire dataflow machinery 
(abstraction functions, merge functions, propagation functions that propagate
dataflow facts across control flow graph nodes) by stating the prophecy 
and history variable specifications over the operational semantics, and 4) 
instead of systematizing dataflow analysis, it relegates dataflow fixed
point computations to an implementation mechanism hidden behind a more abstract
analysis interface. 

PDCE and LCM are classic program analyses and 
transformations~\cite{DBLP:conf/pldi/KnoopRS94,DBLP:journals/toplas/KnoopRS94,DBLP:conf/pldi/KnoopRS92}. 
This previous research formulates the analyses and transformations as multiple 
dataflow analyses over control flow graphs with manual pencil and paper proofs. 
We instead formulate the analysis with prophecy and history variables, which 
are specified over dynamic program traces to promote forward simulation
proofs, relegating dataflow analyses to an implementation mechanism hidden
behind prophecy and history variable abstraction. We also present the first
machine checked proofs of correctness and optimality for these two transformations
and work with a variant of LCM that eliminates a preprocessing phase
that splits critical edges --- our variant instead inserts expression evaluations
on control flow edges (as opposed to only before or after nodes). We present the first 
correctness proof of any kind for this variant. 

The CompCert verified compiler contains an implementation of a
generally standard dataflow analysis framework for supporting
traditional compiler optimizations such as constant propagation
and common subexpression elimination~\cite{BertotGL04}.
The formulation includes the standard dataflow mechanisms --- lattices of dataflow facts,
abstraction functions for mapping register values to lattice values,
and a forward and backward implementation of Kildall's
fixed point algorithm for solving dataflow equations.
Example dataflow domains record when registers contain constant
values (for constant propagation) or the expressions for register
values (for common subexpression elimination). CompCert also 
contains an unverified implementation of lazy code motion, using
a verified validator to prevent unsound transforms from generating
miscompiles~\cite{DBLP:conf/pldi/TristanL09}. Our research differs in that we 
replace the dataflow analysis framework and mechanisms with 
the prophecy and history variable abstraction and our implementation
of lazy code motion is verified --- it comes with correctness and optimality proofs,
not a verified validator for potentially unsound transforms. 

Simulation relations, and techniques for proving that simulation relations exist, have
been extensively explored in the context of establishing simulation relations
between state machines~\cite{LynchV95,LynchDistributedAlgorithms}. The developed theory includes a range
of proof techniques and mechanisms, including forward and backward proof techniques
with refinement mappings, abstraction functions, and abstraction relations.
Prophecy variables were initially developed for the purpose of proving
that implementations satisfy specifications via refinement
mappings with forward simulations, specifically in the case when the
specification makes a choice before the implementation~\cite{AbadiL91}.
The addition of prophecy variables to the framework of refinement mappings
with history variables and forward simulation proofs enabled a completeness
result for the ability to prove trace inclusions of implementations within
specifications~\cite{AbadiL91}. It is, of course, known that backward
simulation is an alternative to forward simulation with prophecy
variables~\cite{LynchV95}. In general, there are a number of alternatives
when choosing a formal framework for proving simulation properties, with the
appropriate framework depending on pragmatic issues
such as the convenience and conceptual difficulty of working with the
concepts in the framework. In general, approaches that reason forward
in time seem to be more attractive and intuitive than approaches that
reason backward against time, as can be seen, for example, in pedagogical
presentations of dataflow analyses, which invariably present forward
analyses first, then backward analyses second as a kind of dual of forward
analyses~\cite{appel2004modern,muchnick1997advanced,DragonBook,cooper2011engineering,kennedy2001optimizing,Aldrich2019Correctness,Aldrich2019Examples}.

Many of the concepts that appear in simulation relation proofs for state
machines also appear in the program verification, dataflow analysis,
and abstract interpretation literature.
For example, history variables were first introduced in the program
verification literature~\cite{OwickiG76}, abstraction functions,
originally introduced in the program verification literature~\cite{Hoare72},
can be seen as a form of refinement mappings,
and program analyses can be seen as establishing a simulation relation between an
abstract interpretation of the program (which plays the role of the
specification) and concrete executions of the program (which play the
role of the implementation)~\cite{CousotC77,CousotC92}. It is also known that, in this context,
backward or reverse simulation relations can be used to establish
the correspondence between backward analyses (which extract
information about the future execution) and program
executions~\cite{CousotC92,SchmidtS98}.

\section{Conclusion}
\label{sec:conclusion}

Dataflow analysis has been a predominant formulation
of program analysis for decades. While it has served the field well, 
its drawbacks include significant mechanism and a mismatch 
between its basic concepts (static control flow graphs, 
lattices, backward and forward analyses) and the basic
concepts (dynamic traces and forward simulation proofs over 
operational semantics) that drive machine checked correctness
and optimality proofs. Prophecy variables (along with history
variables) eliminate the forward/backward distinction in the 
formulation of the analysis. Their close coupling with the 
operational semantics eliminates the dataflow mismatch between
program analysis and forward simulation proofs. With the
recent advent of powerful automated coding agents, the field
will increasingly be driven by automated verification. The
presented prophecy and history variable formulation supports this 
increasingly prominent direction in the field. 

\clearpage 

\section{Repository Road Map}
\label{sec:roadmap}

This paper comes with a repository containing all of the implementation and
proofs for Nexis and the PDCE and LCM transforms. The repository is located at
\url{https://github.com/rinard/Nexis}. All of the code and proofs
in this repository were generated by a coding agent (Claude Code Opus 4.8) 
working with Lean 4.31.0 supervised by the author in VS Code 1.126.0 
working on a MacBook M4 Pro, 12 cores, 24 GB RAM,
macOS 15.1 Darwin 24.1.0, ARM64 ISA. Nexis programs compile to ARM64 ISA.
Figures~\ref{fig:roadmap-intro} through \ref{fig:roadmap-nexis} provide a
roadmap to the repository.

% figures/roadmap-intro.tex --- Repository roadmap for Section 1 (prophecy/history framework).
% \input from the paper body. Self-contained: the link + style macros below are guarded
% (\providecommand), so the roadmap-*.tex files coexist and may be \input in any order.
% ---- link macros ----
\providecommand{\href}[2]{#2}
\providecommand{\nexisroot}{}
\providecommand{\nexisf}[1]{\href{\nexisroot#1}{\texttt{#1}}}
\providecommand{\nexisl}[2]{\href{\nexisroot#1\#L#2}{\texttt{#1:#2}}}
\providecommand{\nexist}[4]{\href{\nexisroot#1\#L#2}{\texttt{#3}}~{\footnotesize[\nexisl{#1}{#2}]}~--- #4}
% ---- styling (idempotent; only xcolor, already loaded by acmart) ----
\definecolor{rmaccent}{HTML}{0F5257}
\definecolor{rmrule}{HTML}{CFCFCF}
\definecolor{rmdim}{HTML}{5F5F5F}
\providecommand{\rmhead}[1]{\noindent{\bfseries #1}\par\vspace{2pt}{\color{rmrule}\rule{\linewidth}{0.7pt}}\par\vspace{3pt}}
\providecommand{\rmnote}[1]{{\footnotesize\itshape\color{rmdim} #1}\par\vspace{3pt}}
\providecommand{\rmgrp}[1]{\vspace{3pt}\par{\bfseries\color{rmaccent}#1}\par\vspace{1pt}}
\providecommand{\rmentry}[1]{\par\hangindent=1.9em\hangafter=1\hspace*{1em}{\color{rmaccent}\footnotesize$\blacktriangleright$}\hspace{0.4em}#1}

\begin{figure*}[t]
\small
\begin{flushleft}
\rmhead{\S\ref{sec:intro}\quad Prophecy \& history variables --- the augmented operational semantics}
\rmnote{The framework of \S\ref{sec:intro}: ghost variables carried alongside base configurations, with
  the analysis defined by constraints on the augmented transitions. Paths relative to the repository root,
  line numbers in parentheses; theorem entries \emph{name}~\texttt{[file:line]}~--- \emph{summary}.}

\rmgrp{Augmented semantics \normalfont(\S\ref{sec:intro}.1 --- ghost-augmented configurations and steps)}
\rmentry{\nexisf{BaseLanguage/Analysis/Augmented.lean} --- \texttt{Aug} (37, a base configuration paired
  with a ghost value), \texttt{AugStep} (44, the augmented transition relation), \texttt{Drives} (56, a
  ghost map $g$ that satisfies the step-rule constraints along every transition).}

\rmgrp{Progress, preservation, bisimulation \normalfont(\S\ref{sec:intro}.2 --- the generic framework)}
\rmentry{\nexist{BaseLanguage/Analysis/Augmented.lean}{61}{progress}{every base transition has a matching
  augmented transition --- prophecy predictions satisfy the downstream checks, so no base step is lost}.}
\rmentry{\nexist{BaseLanguage/Analysis/Augmented.lean}{50}{preservation}{every augmented transition projects
  back to a base transition, preserving the base-variable values}.}
\rmentry{\nexist{BaseLanguage/Analysis/Augmented.lean}{73}{bisim}{progress and preservation together give a
  bisimulation between the base and augmented semantics}.}

\rmgrp{Generated per-analysis instances \normalfont(emitted by \texttt{gen} into \texttt{Seam/}\emph{a}\texttt{/Augmented.lean})}
\rmentry{For every ghost, \texttt{gen} emits its per-step relation, an \texttt{\_drives} validity witness, and
  the framework instances \texttt{\_preservation}/\texttt{\_progress}/\texttt{\_bisim}. PDCE
  (\nexisf{Generated/Seam/pdce/Augmented.lean}): Live ($\pi$) progress \nexisl{Generated/Seam/pdce/Augmented.lean}{21} /
  preservation \nexisl{Generated/Seam/pdce/Augmented.lean}{18}, Sink ($\eta$) progress (43) / preservation (40); LCM
  (\nexisf{Generated/Seam/lcm/Augmented.lean}) covers all six ghosts.}

\rmgrp{Validity \& extremality \normalfont(generated per analysis; machinery in \S\ref{sec:nexis})}
\rmentry{The solver result is a \emph{valid} solution of each ghost's clause predicate and is \emph{extremal}
  (least or greatest) among valid solutions --- exported per analysis in
  \nexisf{Generated/Seam/pdce/ValidExtremal.lean} and \nexisf{Generated/Seam/lcm/ValidExtremal.lean} (bundle theorems
  \texttt{PDCESolve\_valid}, \texttt{LCMSolve\_valid}/\texttt{LCMSolve\_extremal}).}
\end{flushleft}
\caption{Repository roadmap: \S\ref{sec:intro} --- prophecy and history variables as an augmented
  operational semantics (progress, preservation, and the validity/extremality of ghost solutions).}
\label{fig:roadmap-intro}
\end{figure*}

% figures/roadmap-base.tex --- Repository roadmap for Section 3 (base language + verified backend).
% \input from the paper body. Self-contained: the link + style macros below are guarded
% (\providecommand), so the roadmap-*.tex files coexist and may be \input in any order.
% ---- link macros ----
\providecommand{\href}[2]{#2}
\providecommand{\nexisroot}{}
\providecommand{\nexisf}[1]{\href{\nexisroot#1}{\texttt{#1}}}
\providecommand{\nexisl}[2]{\href{\nexisroot#1\#L#2}{\texttt{#1:#2}}}
\providecommand{\nexist}[4]{\href{\nexisroot#1\#L#2}{\texttt{#3}}~{\footnotesize[\nexisl{#1}{#2}]}~--- #4}
% ---- styling (idempotent) ----
\definecolor{rmaccent}{HTML}{0F5257}
\definecolor{rmrule}{HTML}{CFCFCF}
\definecolor{rmdim}{HTML}{5F5F5F}
\providecommand{\rmhead}[1]{\noindent{\bfseries #1}\par\vspace{2pt}{\color{rmrule}\rule{\linewidth}{0.7pt}}\par\vspace{3pt}}
\providecommand{\rmnote}[1]{{\footnotesize\itshape\color{rmdim} #1}\par\vspace{3pt}}
\providecommand{\rmgrp}[1]{\vspace{3pt}\par{\bfseries\color{rmaccent}#1}\par\vspace{1pt}}
\providecommand{\rmentry}[1]{\par\hangindent=1.9em\hangafter=1\hspace*{1em}{\color{rmaccent}\footnotesize$\blacktriangleright$}\hspace{0.4em}#1}

\begin{figure*}[t]
\small
\begin{flushleft}
\rmhead{\S\ref{sec:base-language}\quad Base language --- syntax, reference semantics, and the verified ARM64 backend}
\rmnote{The three-address language of Figures~\ref{fig:ir-syntax}--\ref{fig:ir-semantics} and its verified
  source-to-assembly compiler. Paths relative to the repository root, line numbers in parentheses;
  theorem entries \emph{name}~\texttt{[file:line]}~--- \emph{summary}. The pass pipeline itself is
  Figure~\ref{fig:roadmap-pipeline}.}

\rmgrp{Syntax \& IR \normalfont(Figure~\ref{fig:ir-syntax})}
\rmentry{\nexisf{BaseLanguage/IR/TAC.lean} --- \texttt{Program} (190), \texttt{Cmd} (99), \texttt{Expr} (90),
  \texttt{Config} (258), \texttt{Store} (234); \nexisf{BaseLanguage/IR/Cfg.lean} --- \texttt{succList} (24, the
  successor function $\mathsf{succ}(n)$).}

\rmgrp{Reference semantics \normalfont(Figures~\ref{fig:ir-eval}--\ref{fig:ir-semantics})}
\rmentry{\nexisf{BaseLanguage/IR/TAC.lean} --- \texttt{eval} (252, expression evaluation), \texttt{Step} (264,
  the transition relation $\langle n,\sigma\rangle\to\langle n',\sigma'\rangle$), \texttt{run} (359, functional
  reference interpreter), \texttt{WellFormed} (205), \texttt{Status} (304).}
\rmentry{\nexisf{BaseLanguage/Behavior/Outcomes.lean} --- the three outcomes \texttt{Halts} (117),
  \texttt{Faults} (120), \texttt{Diverges} (123).}

\rmgrp{High-level language \& lowering}
\rmentry{\nexisf{BaseLanguage/Frontend/Ast.lean} --- \texttt{Stmt} (30, structured control flow);
  \nexisf{BaseLanguage/Pass/AstToTac.lean} --- \texttt{lower} (131, to three-address code).}
\rmentry{\nexist{BaseLanguage/Pass/Correctness/AstToTacCorrect.lean}{648}{lower\_correct}{lowering preserves
  the observable result of the high-level program}.}

\rmgrp{Verified ARM64 backend}
\rmentry{\nexisf{BaseLanguage/Backend/Asm.lean} --- \texttt{Cmd} (99, the ARM64 machine model);
  \nexisf{BaseLanguage/Backend/TacToAsm.lean} --- \texttt{codegen} (113).}
\rmentry{\nexist{BaseLanguage/Backend/Correctness/CodegenForward.lean}{346}{codegen\_simulates}{the emitted
  assembly simulates the IR reference semantics}.}
\rmentry{\nexist{BaseLanguage/Pass/Correctness/AstToTacCorrect.lean}{667}{ast\_to\_asm}{source to assembly,
  end to end}; \nexist{BaseLanguage/Pass/Correctness/PipelineToAsm.lean}{71}{pipeline\_to\_asm}{the full
  optimizing pipeline preserves the observable result}.}

\rmgrp{Text interfaces \normalfont(unverified: parser and printers)}
\rmentry{\nexisf{BaseLanguage/Frontend/TextToAst.lean} --- \texttt{parse} (196);
  \nexisf{BaseLanguage/Backend/AsmToText.lean} --- \texttt{emitText} (92);
  \nexisf{BaseLanguage/Frontend/AstToText.lean} --- \texttt{emitText} (87).}
\end{flushleft}
\caption{Repository roadmap: \S\ref{sec:base-language} --- the base language, its reference semantics, and
  the verified source-to-ARM64 compiler (code generation and end-to-end correctness).}
\label{fig:roadmap-base}
\end{figure*}

% figures/roadmap-pipeline.tex --- Repository roadmap for Section 3 (the verified pass pipeline + tests).
% \input from the paper body. Self-contained: the link + style macros below are guarded
% (\providecommand), so the roadmap-*.tex files coexist and may be \input in any order.
% ---- link macros ----
\providecommand{\href}[2]{#2}
\providecommand{\nexisroot}{}
\providecommand{\nexisf}[1]{\href{\nexisroot#1}{\texttt{#1}}}
\providecommand{\nexisl}[2]{\href{\nexisroot#1\#L#2}{\texttt{#1:#2}}}
\providecommand{\nexist}[4]{\href{\nexisroot#1\#L#2}{\texttt{#3}}~{\footnotesize[\nexisl{#1}{#2}]}~--- #4}
% ---- styling (idempotent) ----
\definecolor{rmaccent}{HTML}{0F5257}
\definecolor{rmrule}{HTML}{CFCFCF}
\definecolor{rmdim}{HTML}{5F5F5F}
\providecommand{\rmhead}[1]{\noindent{\bfseries #1}\par\vspace{2pt}{\color{rmrule}\rule{\linewidth}{0.7pt}}\par\vspace{3pt}}
\providecommand{\rmnote}[1]{{\footnotesize\itshape\color{rmdim} #1}\par\vspace{3pt}}
\providecommand{\rmgrp}[1]{\vspace{3pt}\par{\bfseries\color{rmaccent}#1}\par\vspace{1pt}}
\providecommand{\rmentry}[1]{\par\hangindent=1.9em\hangafter=1\hspace*{1em}{\color{rmaccent}\footnotesize$\blacktriangleright$}\hspace{0.4em}#1}

\begin{figure*}[t]
\small
\begin{flushleft}
\rmhead{\S\ref{sec:base-language}\quad Compiler pipeline --- the verified pass sequence and test programs}
\rmnote{The passes of \S\ref{sec:base-language}, in driver order
  $\mathsf{codegen}\circ\mathsf{cleanup}\circ\mathsf{PDCE}\circ\mathsf{LCM}\circ\mathsf{normalize}\circ
  \mathsf{iterateOpt}\circ\mathsf{normalize}\circ\mathsf{peephole}\circ\mathsf{lower}$ (LCM/PDCE detailed in
  Figures~\ref{fig:roadmap-pdce}--\ref{fig:roadmap-lcm}). Each pass preserves halting, faulting, and
  divergence; line numbers in parentheses.}

\rmgrp{Peephole \normalfont(constant folding + algebraic identities)}
\rmentry{\nexisf{BaseLanguage/Peephole/Simplify.lean} (expression rewrites); \nexisf{BaseLanguage/Peephole/Pass.lean}
  --- \texttt{peephole} (32), \texttt{peephole\_preserves\_halt} (94), \texttt{\_diverge} (117), \texttt{\_faultSteps} (123).}

\rmgrp{Normalize \normalfont(the \texttt{WellNormalized} foundation)}
\rmentry{\nexisf{BaseLanguage/Normalize/Normalize.lean} --- \texttt{normalize} (24),
  \nexist{BaseLanguage/Normalize/Normalize.lean}{40}{normalize\_wellNormalized}{enforces the six structural
  invariants (in-range successors, no self-reads, no duplicate/entry successors, reachable, noop entry)}.}
\rmentry{\nexist{BaseLanguage/Normalize/FromLower.lean}{337}{normalize\_lower\_wellNormalized}{the turnkey
  $\mathsf{normalize}\circ\mathsf{lower}$ is well-normalized unconditionally};
  \nexisf{BaseLanguage/Normalize/SelfRead.lean} --- \texttt{normalizeSelfRead} (50), \texttt{normalize\_noSelfRead} (152).}

\rmgrp{Iterated constant propagation \normalfont(run to a fixed point)}
\rmentry{\nexisf{BaseLanguage/Pass/Optimize.lean} --- \texttt{iterateOpt} (81),
  \texttt{iterateOpt\_preserves\_halt} (102)/\texttt{\_faults} (117)/\texttt{\_diverges} (131).}
\rmentry{\emph{Constant fold:} \nexisf{BaseLanguage/Pass/ConstFold.lean} --- \texttt{run} (74),
  \texttt{preserves\_halt} (134)/\texttt{\_faults} (141)/\texttt{\_diverges} (152).}
\rmentry{\emph{Branch fold:} \nexisf{BaseLanguage/Pass/BranchFold.lean} --- \texttt{run} (134),
  \texttt{preserves\_halt} (206)/\texttt{\_faults} (213)/\texttt{\_diverges} (224).}
\rmentry{\emph{Unreachable-code elimination:} \nexisf{BaseLanguage/Pass/UCE.lean} --- \texttt{run} (94),
  \texttt{preserves\_halt} (152)/\texttt{\_faults} (162)/\texttt{\_diverges} (180).}

\rmgrp{Cleanup \& driver}
\rmentry{\nexisf{BaseLanguage/Pass/Cleanup.lean} --- \texttt{cleanup} (72, remove superfluous noops);
  \texttt{cleanup\_preserves\_halt} (\nexisl{BaseLanguage/Pass/CleanupCorrect.lean}{421})/\texttt{\_faults} (437)/\texttt{\_diverges} (507).}
\rmentry{\nexisf{Seam/compile/CompileCorrect.lean} --- \texttt{optProvider} (26, wires the analyses into the
  driver), \nexist{Seam/compile/CompileCorrect.lean}{53}{main\_compile\_correct}{the whole compiler preserves
  observable behavior}; CLI \nexisf{Main.lean} --- \texttt{compile} (81), \texttt{showOpt} (86), \texttt{main} (111).}

\rmgrp{Test programs}
\rmentry{\nexisf{test/Tests.lean} --- \texttt{computationCases} (166, algorithms with known results),
  \texttt{optCases} (203, optimizer demos); runnable inputs in \nexisf{examples/computations/} and
  \nexisf{examples/scale/}, with per-stage dumps (lowered IR / optimized IR / assembly) in \nexisf{examples/opt/dumps/}.}
\end{flushleft}
\caption{Repository roadmap: \S\ref{sec:base-language} --- the verified compiler pass pipeline
  (peephole, normalization, iterated constant propagation, cleanup) and the test programs.}
\label{fig:roadmap-pipeline}
\end{figure*}

% figures/roadmap-pdce.tex --- Repository roadmap for Section 4 (Partial Dead Code Elimination).
% \input from the paper body. Self-contained: the link + style macros below are guarded
% (\providecommand), so the roadmap-*.tex files coexist and may be \input in any order.
% ---- link macros ----
\providecommand{\href}[2]{#2}
\providecommand{\nexisroot}{}
\providecommand{\nexisf}[1]{\href{\nexisroot#1}{\texttt{#1}}}
\providecommand{\nexisl}[2]{\href{\nexisroot#1\#L#2}{\texttt{#1:#2}}}
\providecommand{\nexist}[4]{\href{\nexisroot#1\#L#2}{\texttt{#3}}~{\footnotesize[\nexisl{#1}{#2}]}~--- #4}
% ---- styling (idempotent) ----
\definecolor{rmaccent}{HTML}{0F5257}
\definecolor{rmrule}{HTML}{CFCFCF}
\definecolor{rmdim}{HTML}{5F5F5F}
\providecommand{\rmhead}[1]{\noindent{\bfseries #1}\par\vspace{2pt}{\color{rmrule}\rule{\linewidth}{0.7pt}}\par\vspace{3pt}}
\providecommand{\rmnote}[1]{{\footnotesize\itshape\color{rmdim} #1}\par\vspace{3pt}}
\providecommand{\rmgrp}[1]{\vspace{3pt}\par{\bfseries\color{rmaccent}#1}\par\vspace{1pt}}
\providecommand{\rmentry}[1]{\par\hangindent=1.9em\hangafter=1\hspace*{1em}{\color{rmaccent}\footnotesize$\blacktriangleright$}\hspace{0.4em}#1}

\begin{figure*}[t]
\small
\begin{flushleft}
\rmhead{\S\ref{sec:pdce}\quad Partial dead code elimination --- analysis, transform, correctness, optimality}
\rmnote{The prophecy variable $\pi$ (live) and history variable $\eta$ (moving assignments) of
  Figures~\ref{fig:pdce-spec}--\ref{fig:pdce-locals}, the transform of Figure~\ref{fig:pdce-insert}, and the
  proofs of \S\ref{sec:pdce}. Line numbers in parentheses; theorem entries \emph{name}~\texttt{[file:line]}~--- \emph{summary}.}

\rmgrp{Analysis \normalfont(Figures~\ref{fig:pdce-spec}--\ref{fig:pdce-locals}; ghost specification + node locals)}
\rmentry{\nexisf{analyses/pdce/Pdce.gsl} (the two-ghost specification);
  \nexisf{analyses/pdce/PdceDefs.lean} --- \texttt{born} (29), \texttt{pass} (61), \texttt{defVars} (48),
  \texttt{rhsVars} (43), \texttt{condVars} (52), \texttt{liveSeed} (74).}
\rmentry{Generated: \nexisf{Generated/Solver/pdce/Solve.lean} (dataflow solve) and \nexisf{Generated/Seam/pdce/ValidExtremal.lean}
  (clause predicates \texttt{Live}/\texttt{Sink}; bundle \texttt{PDCESolve\_valid}).}

\rmgrp{Transform \normalfont(Figure~\ref{fig:pdce-insert})}
\rmentry{\nexisf{BaseLanguage/PDCE/Transform.lean} --- \texttt{transform} (98, noop-out then reinsert),
  \texttt{delayedExit} (32); block-layout scaffolding in \nexisf{BaseLanguage/PDCE/Layout.lean}.}

\rmgrp{Correctness \normalfont(\S\ref{sec:pdce}.3; forward simulation, valid $G$ suffices)}
\rmentry{\nexist{BaseLanguage/PDCE/Correctness.lean}{1150}{transform\_preserves\_halt}{if the original halts
  the transform halts with the same observables (via the \texttt{match\_step} simulation)}.}
\rmentry{\nexist{BaseLanguage/PDCE/Divergence.lean}{24}{transform\_preserves\_diverges}{the transform
  preserves divergence}.}

\rmgrp{Optimality \normalfont(\S\ref{subsec:pdce-opt}; extremal $G$ vs.\ arbitrary valid $G'$)}
\rmentry{\emph{Assignment optimality:} \nexist{BaseLanguage/PDCE/ExecCount/Headline.lean}{989}{transform\_execCount\_le\_safe}{no
  assignment runs more often than any safe placement}; two-program forms
  \texttt{matCount\_le\_any} (\nexisl{BaseLanguage/PDCE/ExecCountTwoProgram.lean}{180}),
  \texttt{transform\_execCountOrig\_le\_any} (194).}
\rmentry{\emph{Live-range optimality:} \nexist{BaseLanguage/PDCE/Optimality.lean}{175}{pathVarLiveRange\_le}{extremal
  placement minimizes each variable's live range}; two-program forms
  \texttt{transform\_regOccOrig\_le\_any} (\nexisl{BaseLanguage/PDCE/LiveRangeTwoProgram.lean}{91}),
  \texttt{transform\_regOccExtOrig\_le\_any} (211); reads-fixed form
  \texttt{transform\_honestLiveRange\_le} (\nexisl{BaseLanguage/PDCE/OperationalLiveRange.lean}{126}).}
\end{flushleft}
\caption{Repository roadmap: \S\ref{sec:pdce} --- PDCE analysis (prophecy/history ghosts), transform, and the
  machine-checked correctness and optimality proofs.}
\label{fig:roadmap-pdce}
\end{figure*}

% figures/roadmap-lcm.tex --- Repository roadmap for Section 5 (Lazy Code Motion) + the simplified variant.
% \input from the paper body. Self-contained: the link + style macros below are guarded
% (\providecommand), so the roadmap-*.tex files coexist and may be \input in any order.
% ---- link macros ----
\providecommand{\href}[2]{#2}
\providecommand{\nexisroot}{}
\providecommand{\nexisf}[1]{\href{\nexisroot#1}{\texttt{#1}}}
\providecommand{\nexisl}[2]{\href{\nexisroot#1\#L#2}{\texttt{#1:#2}}}
\providecommand{\nexist}[4]{\href{\nexisroot#1\#L#2}{\texttt{#3}}~{\footnotesize[\nexisl{#1}{#2}]}~--- #4}
% ---- styling (idempotent) ----
\definecolor{rmaccent}{HTML}{0F5257}
\definecolor{rmrule}{HTML}{CFCFCF}
\definecolor{rmdim}{HTML}{5F5F5F}
\providecommand{\rmhead}[1]{\noindent{\bfseries #1}\par\vspace{2pt}{\color{rmrule}\rule{\linewidth}{0.7pt}}\par\vspace{3pt}}
\providecommand{\rmnote}[1]{{\footnotesize\itshape\color{rmdim} #1}\par\vspace{3pt}}
\providecommand{\rmgrp}[1]{\vspace{3pt}\par{\bfseries\color{rmaccent}#1}\par\vspace{1pt}}
\providecommand{\rmentry}[1]{\par\hangindent=1.9em\hangafter=1\hspace*{1em}{\color{rmaccent}\footnotesize$\blacktriangleright$}\hspace{0.4em}#1}

\begin{figure*}[t]
\small
\begin{flushleft}
\rmhead{\S\ref{sec:lcm}\quad Lazy code motion --- six interdependent ghosts, transform, correctness, optimality}
\rmnote{The six ghost variables ($\pi_a,\eta_a,\eta_p,\tau_p,\pi_u,\tau_u$) of
  Figures~\ref{fig:lcm-spec}--\ref{fig:lcm-locals}, the critical-edge transform of Figure~\ref{fig:lcm-insert},
  and the proofs of \S\ref{sec:lcm}. Line numbers in parentheses; theorem entries \emph{name}~\texttt{[file:line]}~--- \emph{summary}.}

\rmgrp{Analysis \normalfont(Figures~\ref{fig:lcm-spec}--\ref{fig:lcm-locals}; six ghosts + node locals)}
\rmentry{\nexisf{analyses/lcm/Lcm.gsl} (the six-ghost specification);
  \nexisf{analyses/lcm/LcmDefs.lean} --- \texttt{ue} (24), \texttt{de} (40), \texttt{pass} (39),
  \texttt{earliest} (50), \texttt{latestNode} (55), \texttt{latestEdge} (57).}
\rmentry{Generated: \nexisf{Generated/Solver/lcm/Solve.lean} (dependence-ordered solve) and
  \nexisf{Generated/Seam/lcm/ValidExtremal.lean} (clause predicates \texttt{Anticipated}/\texttt{Available}/\texttt{Postponable}/\texttt{Used}
  plus the two transfer variables; bundle \texttt{LCMSolve\_valid} (208), \texttt{LCMSolve\_extremal} (211)).}

\rmgrp{Transform \normalfont(Figure~\ref{fig:lcm-insert}; edge insertions, no critical-edge splitting)}
\rmentry{\nexisf{BaseLanguage/LCM/Transform.lean} --- \texttt{transform} (205), \texttt{insertBefore} (96),
  \texttt{insertEdge} (114), \texttt{recoverable} (127); block layout in \nexisf{BaseLanguage/LCM/Layout.lean}.}
\rmentry{Self-read normalization (required precondition): \nexisf{BaseLanguage/Normalize/SelfRead.lean} ---
  \texttt{normalizeSelfRead} (50), \texttt{normalize\_noSelfRead} (152).}

\rmgrp{Correctness \normalfont(\S\ref{sec:lcm}.3; forward simulation, requires extremal $G$)}
\rmentry{\nexist{BaseLanguage/LCM/Correctness/MatchStep.lean}{707}{transform\_preserves\_halt}{if the original halts the
  transform halts with the same observables (the \texttt{Match}/\texttt{Cov} invariant)}.}
\rmentry{\nexist{BaseLanguage/LCM/Correctness/FaultPreservation.lean}{370}{transform\_preserves\_faulting}{the transform
  preserves faults}.}

\rmgrp{Optimality \normalfont(\S\ref{sec:lcm}.4)}
\rmentry{\emph{Computational optimality:} \nexist{BaseLanguage/LCM/EvalCountHeadline.lean}{381}{transform\_evalCount\_le\_safe}{on
  every halting trace, each expression is evaluated no more often than under any safe placement}.}
\rmentry{\emph{Lifetime optimality:} \nexist{BaseLanguage/LCM/Optimality.lean}{240}{pathLiveLen\_le}{among
  computationally optimal placements, extremal $\eta_p$ minimizes temporary live ranges} (on
  \texttt{liveRegion\_minimal}, \nexisl{BaseLanguage/LCM/Optimality.lean}{52}); supporting evaluation-count
  development in \nexisf{BaseLanguage/LCM/EvalCount.lean}, \nexisf{BaseLanguage/LCM/EvalCountPlace.lean}.}

\rmgrp{Simplified variant \normalfont(\S\ref{sec:lcm-insert-basic}, Figure~\ref{fig:lcm-insert-basic}; drops $\pi_u,\tau_u$, allows isolated computations)}
\rmentry{\nexisf{BaseLanguage/LCM/BasicCorrect.lean} --- \texttt{transform\_preserves\_halt\_basic} (203),
  \texttt{transform\_preserves\_faulting\_basic} (487).}
\end{flushleft}
\caption{Repository roadmap: \S\ref{sec:lcm} --- LCM analysis (six ghosts), the critical-edge transform,
  correctness and optimality proofs, and the simplified variant of \S\ref{sec:lcm-insert-basic}.}
\label{fig:roadmap-lcm}
\end{figure*}

% figures/roadmap-nexis.tex --- Repository roadmap for Section 6 (Nexis: language, generator, solvers).
% \input from the paper body. Self-contained: the link + style macros below are guarded
% (\providecommand), so the roadmap-*.tex files coexist and may be \input in any order.
% ---- link macros ----
\providecommand{\href}[2]{#2}
\providecommand{\nexisroot}{}
\providecommand{\nexisf}[1]{\href{\nexisroot#1}{\texttt{#1}}}
\providecommand{\nexisl}[2]{\href{\nexisroot#1\#L#2}{\texttt{#1:#2}}}
\providecommand{\nexist}[4]{\href{\nexisroot#1\#L#2}{\texttt{#3}}~{\footnotesize[\nexisl{#1}{#2}]}~--- #4}
% ---- styling (idempotent) ----
\definecolor{rmaccent}{HTML}{0F5257}
\definecolor{rmrule}{HTML}{CFCFCF}
\definecolor{rmdim}{HTML}{5F5F5F}
\providecommand{\rmhead}[1]{\noindent{\bfseries #1}\par\vspace{2pt}{\color{rmrule}\rule{\linewidth}{0.7pt}}\par\vspace{3pt}}
\providecommand{\rmnote}[1]{{\footnotesize\itshape\color{rmdim} #1}\par\vspace{3pt}}
\providecommand{\rmgrp}[1]{\vspace{3pt}\par{\bfseries\color{rmaccent}#1}\par\vspace{1pt}}
\providecommand{\rmentry}[1]{\par\hangindent=1.9em\hangafter=1\hspace*{1em}{\color{rmaccent}\footnotesize$\blacktriangleright$}\hspace{0.4em}#1}

\begin{figure*}[t]
\small
\begin{flushleft}
\rmhead{\S\ref{sec:nexis}\quad Nexis --- ghost-spec language, generator, and verified dataflow solvers}
\rmnote{The specification language, the (unverified) generator that lowers it, and the verified solver suite
  behind it. Line numbers in parentheses; theorem entries \emph{name}~\texttt{[file:line]}~--- \emph{summary}.}

\rmgrp{Ghost-spec language \& Defs files \normalfont(\S\ref{sec:nexis}.1--\S\ref{sec:nexis}.3)}
\rmentry{Language reference \nexisf{GHOST-SPEC-LANGUAGE.md} and frozen grammar \nexisf{docs/GENERAL-PIPELINE-EBNF.md}
  (\texttt{predict}/\texttt{check}/\texttt{update}/\texttt{always}/\texttt{seed}/\texttt{within}); node-local and
  placement functions authored per analysis in \nexisf{analyses/pdce/PdceDefs.lean}, \nexisf{analyses/lcm/LcmDefs.lean}.}

\rmgrp{Generator \normalfont(\S\ref{sec:nexis}.2; unverified translator, pure data --- no reflection)}
\rmentry{\nexisf{GenGeneral/} --- \texttt{parseGslBlocks} (\nexisl{GenGeneral/Parser.lean}{399}),
  \texttt{lowerGsl} (\nexisl{GenGeneral/Lower.lean}{381}), \texttt{selectAll} (\nexisl{GenGeneral/Select.lean}{53},
  infers each ghost's quadrant), \texttt{emitSolve} (\nexisl{GenGeneral/Emit.lean}{372}),
  \texttt{emitValidExtremal} (851), \texttt{emitAugmented} (840), \texttt{manifest} (\nexisl{GenGeneral/Driver.lean}{49});
  driven by \nexisf{Gen.lean} (\texttt{lake exe gen}).}
\rmentry{Three generated files per analysis: \nexisf{Generated/Solver/}\emph{a}\nexisf{/Solve.lean} (transfer functions +
  solver invocation), \nexisf{Generated/Seam/}\emph{a}\nexisf{/ValidExtremal.lean} (clause restatements, validity/
  extremality, and the ghost bundle the transforms consume), and \nexisf{Generated/Seam/}\emph{a}\nexisf{/Augmented.lean}
  (the per-ghost augmented-semantics instances --- Progress/Preservation of \S\ref{sec:intro}, Fig.~\ref{fig:roadmap-intro}).}

\rmgrp{Dataflow backend --- universe \& four quadrants \normalfont(\S\ref{sec:nexis}.4.1--\S\ref{sec:nexis}.4.2)}
\rmentry{Finite sets as fixed-width bitvectors: \nexisf{Solver/Impl/Core.lean} --- \texttt{ESet} (28, \texttt{BitVec n}).}
\rmentry{The four solve engines (direction $\times$ confluence), \nexisf{Solver/Impl/Term.lean} ---
  \texttt{solveMTC} (203, fwd$\cdot$must), \texttt{solveMTCMF} (770, fwd$\cdot$may), \texttt{solveMTCBM} (590,
  bwd$\cdot$must), \texttt{solveMTCB} (384, bwd$\cdot$may); transfer-variable confluence
  \texttt{resMeet}/\texttt{resJoin} (\nexisl{Solver/Impl/Transfer.lean}{46}).}

\rmgrp{Solvers \& proofs \normalfont(\S\ref{sec:nexis}.4.4)}
\rmentry{Efficient worklist vs.\ naive reference engine: \nexisf{Solver/Impl/Worklist.lean} --- \texttt{solveWL} (75),
  \nexist{Solver/Impl/Worklist.lean}{715}{solveWL\_eq\_must}{the worklist computes the same result as the naive
  engine (dually \texttt{solveWL\_eq\_may}, 737)}; reference \nexisf{Solver/Impl/Engine.lean}.}
\rmentry{The seam contract + lifted correctness: \nexisf{Solver/Quadrant.lean} pins the eight exported symbols
  \texttt{resMTC}/\texttt{resMTCMF}/\texttt{resMTCBM}/\texttt{resMTCB} and their \texttt{\_correct} theorems
  (valid + extremal, over finite sets and the operational semantics).}
\rmentry{Axiom discipline: \nexisf{BaseLanguage/Meta/AxiomCheck.lean} --- \texttt{\#assert\_clean\_axioms} (18)
  fails the build unless a constant depends only on \texttt{propext}, \texttt{Classical.choice}, \texttt{Quot.sound}
  (rejects \texttt{sorryAx}); invoked on every emitted validity/extremality theorem.}
\end{flushleft}
\caption{Repository roadmap: \S\ref{sec:nexis} --- the Nexis ghost-spec language, the generator that lowers it
  to per-analysis Lean, and the verified four-quadrant dataflow solver suite.}
\label{fig:roadmap-nexis}
\end{figure*}

\clearpage

% ===== Fig. 6' — LCM hoisting placement, BASIC variant (isolated insertions permitted) =====
% Drops the Used analysis (pi_u / tau_u): inserts at every latest node/edge, replaces every
% numbered computation. Correct but not lifetime-optimal (may emit isolated temps).
\begin{figure}[thbp]
\small
\begin{mathpar}
\inferrule*[left=\textsc{Ins-entry}]
  {e \in \mathsf{insertBefore}(n)}
  {t_e := e\ \in\ \mathsf{entry}(n)}
\\
\inferrule*[left=\textsc{Ins-edge}]
  {m \in \mathsf{succ}(n) \\ e \in \mathsf{insertEdge}(n,m)}
  {t_e := e\ \in\ \mathsf{edge}(n,m)}
\\
\inferrule*[left=\textsc{Replace}]
  {\mathsf{cmd}(n) = (x := e) \\ \mathsf{numbered}(e)}
  {\mathsf{ctrl}(n) =\ x := t_e}
\\
\inferrule*[left=\textsc{Keep}]
  {\mathsf{cmd}(n) = (x := e) \\ \neg\,\mathsf{numbered}(e)}
  {\mathsf{ctrl}(n) =\ x := e}
\end{mathpar}

\medskip
\noindent$
\begin{array}{@{}l@{\ }c@{\ }l@{}}
\mathsf{insertEdge}(n,m)  & = & \mathsf{latestEdge}(\pi_a,\eta_a,\eta_p)(n,m)\\
\mathsf{insertBefore}(n)  & = & \mathsf{latestNode}(\eta_p,\tau_p)(n) \setminus \textstyle\bigcup_{m \in \mathsf{succ}(n)} \mathsf{latestEdge}(\pi_a,\eta_a,\eta_p)(n,m)
\end{array}$
\caption{LCM expression evaluations inserted before nodes $\textsc{Ins-entry}$ and on edges $\textsc{Ins-edge}$, simplified variant.}
\label{fig:lcm-insert-basic}
\end{figure}

\section{Lazy Code Motion, Simplified Variant}
\label{sec:lcm-insert-basic}

Figure~\ref{fig:lcm-insert-basic} presents a simplified version of LCM
that discards $\pi_u$ and $\tau_u$ and allows the insertion of
isolated computations $t_e := e; x:=t_e$, where $x:=t_e$ is the only use of $t_e$.

\clearpage

% match_step and invariant Match in ProphecyVariable/BaseLanguage/PDCE/Correctness.lean

% The ultimate goal of this research is to replace dataflow analysis as a top level abstraction in program analysis and relegate it to an implementation mechanism for a cleaner abstraction. 

%% -----------------------------------------------------------------
%% Acknowledgments (remove/comment out for the anonymized submission;
%% acmart also auto-suppresses this environment when `anonymous` is set)
%% -----------------------------------------------------------------
% \begin{acks}
% We thank \ldots for helpful feedback. This work was supported by \ldots
% \end{acks}

%% -----------------------------------------------------------------
%% Bibliography
%% -----------------------------------------------------------------
%% acmart defaults to the ACM Reference Format via natbib + acmart.bst.
%% Put your .bib file(s) in the same directory and list them below.
\bibliographystyle{ACM-Reference-Format}
\bibliography{paper}

%% -----------------------------------------------------------------
%% Appendix (optional; POPL allows a supplementary appendix that
%% reviewers are not obligated to read)
%% -----------------------------------------------------------------
% \appendix
% \section{Additional Proofs}
% Full proofs elided from the main text for space.

\end{document}